\documentclass[12pt,preprint]{aastex}

\slugcomment{}

\shorttitle{Mid-IR spectra of ULIRGs}
\shortauthors{Farrah et al}

\begin{document}

\title{High resolution mid-infrared spectroscopy of ultraluminous infrared galaxies}

\author{D. Farrah\altaffilmark{1}}
\author{J. Bernard-Salas\altaffilmark{1}}
\author{H. W. W. Spoon\altaffilmark{1}}
\author{B. T. Soifer\altaffilmark{2}}
\author{L. Armus\altaffilmark{2}}
\author{B. Brandl\altaffilmark{3}}
\author{V. Charmandaris\altaffilmark{4,5}}
\author{V. Desai\altaffilmark{2}}
\author{S. Higdon\altaffilmark{6}}
\author{D. Devost\altaffilmark{1}}
\author{J. Houck\altaffilmark{1}}

\altaffiltext{1}{Department of Astronomy, Cornell University, Ithaca, NY 14853, USA}
\altaffiltext{2}{Spitzer Science Center, 1200 East California Boulevard, Pasadena, CA 91125, USA}
\altaffiltext{3}{Sterrewacht Leiden, Leiden University, P.O. Box 9513, Niels Bohrweg 2, 2300 RA Leiden, The Netherlands}
\altaffiltext{4}{Department of Physics, University of Crete, GR-71003 Heraklion, Greece}
\altaffiltext{5}{IESL/Foundation for Research and Technology - Hellas, GR-71110, Heraklion, Greece and Chercheur Associ\'e, Observatoire de Paris, F-75014, Paris, France}
\altaffiltext{6}{Physics Dept, Georgia Southern University, Statesboro, GA 30460, USA}

\begin{abstract}
We present $R\sim600$, 10$\mu$m - 37$\mu$m spectra of 53 Ultraluminous Infrared Galaxies (ULIRGs)  at $z<0.32$, taken
using the Infrared Spectrograph on board the {\it Spitzer} space telescope. All of the spectra show various fine
structure emission lines of Neon, Oxygen, Sulfur, Silicon and Argon, as well as one or more molecular Hydrogen lines.
Some objects also show emission lines from Chlorine, Iron, Phosphorous and atomic Hydrogen, as well as absorption features arising from
C$_{2}$H$_{2}$, HCN, and OH$^{-}$. We employ diagnostics based on the fine-structure emission lines, as well as the
equivalent widths and luminosities of polycyclic aromatic hydrocarbon (PAH) features and the strength of the 9.7$\mu$m
silicate absorption feature ($S_{sil}$), to explore the power source behind the infrared emission in ULIRGs. We show
that the infrared emission from the majority of ULIRGs is powered mostly by star formation, with only $\sim$20\% of ULIRGs hosting an AGN with
a comparable or greater IR luminosity than the starburst. The detection of the [NeV]$\lambda$14.32 line in just under
half the sample however implies that an AGN makes a significant contribution to the mid-IR flux in $\sim42\%$ of ULIRGs.
The fine structure line ratios, luminosities and PAH EWs of our sample are consistent
with the starbursts and AGN in ULIRGs being more extincted, and for the starbursts more compact, versions of those in lower luminosity systems.
The excitations and electron densities in the narrow-line regions of ULIRGs appear comparable to those of lower
luminosity ($10^{10}<$L$_{ir}$(L$_{\odot})<10^{11.5}$) starbursts, though there is evidence that the NLR gas in ULIRGs
is more dense. We show that the combined luminosity of the [NeII]$\lambda$12.81 and [NeIII]$\lambda$15.56 lines correlates with with
both infrared luminosity and luminosity of  the 6.2$\mu$m and 11.2$\mu$m PAH features in ULIRGs, and use this to derive
a calibration between PAH luminosity and star formation rate. Finally, we show that those ULIRGs with
$0.8\lesssim S_{sil}\lesssim 2.4$ are likely to be powered mainly by star formation, but that those with
$S_{sil}\lesssim0.8$, and possibly those with $S_{sil}\gtrsim 2.4$, contain an IR-luminous AGN.

\end{abstract}

%% Keywords should appear after the \end{abstract} command. The uncommented
%% example has been keyed in ApJ style. See the instructions to authors
%% for the journal to which you are submitting your paper to determine
%% what keyword punctuation is appropriate.


\section{Introduction}
Ultraluminous Infrared Galaxies (ULIRGs, those objects with 1-1000$\mu$m luminosities in excess of $10^{12}L_{\odot}$) were 
first discovered in the 1970s \citep{rie}. Since then, they have fascinated astronomers with their unique and extraordinary properties, 
and infuriated them with their singularly opaque natures, almost in equal measure. 

In the local Universe ULIRGs are a rare, if interesting oddity, with only fifty or so examples known at $z\lesssim0.1$. First 
uncovered in significant numbers by surveys with the Infrared Astronomical Satellite (IRAS, \citealt{soi1,hou85}), work 
focused on determining the power source behind their colossal infrared emission. This initially provoked heated debate 
between a 'starburst' camp and an 'AGN' camp, a debate that has not yet entirely cooled. Early studies showed that some ULIRG 
optical spectra resembled those of starburst galaxies \citep{jow}, whereas others contained emission lines characteristic of 
Seyferts \citep{san2}. Radio observations showed direct evidence for starbursts in some ULIRGs \citep{con1,smi1} and AGN 
in others \citep{lon03,nag}. Over the last decade or so however a consensus has started to emerge; local ULIRGs are 
likely to be `composite' objects, with most powered mainly by a starburst, but with a significant fraction also containing an IR-luminous 
AGN. This is suggested from several lines of evidence, including optical/UV spectroscopy \citep{vei,vei2,lip,far05}, 
mid-infared spectroscopy \citep{lut96,gen,lut98,rig,gen00,tra01}, modelling of their 1-1000$\mu$m spectral energy distributions \citep{kla01,far4}, and 
X-ray observations \citep{fra03,pta03}, though some recent studies suggest a significantly greater average AGN contribution \citep{ima07}. 
Local ULIRGs are also associated almost exclusively with galaxy mergers \citep{far1,bus2002,vei3}, and may 
be involved to some degree in triggering QSOs \citep{san2,tac02,kaw06,zau07}. Excellent reviews of the properties of ULIRGs can be 
found in \citet{san96}, and more recently in \citet{lfs06}. 

Their rarity in the local Universe compared to lower luminosity systems initially led astronomers to believe that ULIRGs did not play a 
fundamental role in galaxy formation processes, however this perception changed abruptly 
when it was realised that ULIRGs were vastly more numerous at high redshift. First hinted at by 
spectroscopic followup of IRAS surveys \citep{hac87,lon90,sau}, which showed strong evolution in the ULIRG luminosity function with redshift, 
and from the discovery of a remarkably high cosmic infrared background by COBE \citep{pug}, this was confirmed by surveys with ISO 
\citep{rr97,lev98,dol,ver05}, which found a large population of ULIRGs up to $z\sim1.5$, and thrown into sharp relief by sub-mm surveys 
\citep{hug,eal,bor,cop}, which showed that there were several hundred ULIRGs {\it per square degree} at $z\gtrsim1$. 
Though obviously much harder to study, these distant ULIRGs seem superficially similar to their low redshift counterparts in that they 
appear to be powered by both starburst and AGN activity \citep{far3,sma03,sma04,ale05,tak06,val07}, and are probably mergers 
\citep{far2,cha03}. Their properties may make them important tools in understanding 
the global evolution of galaxies and large-scale structures; their rapid star formation rates and comoving number densities make 
them strong candidates for being the rapid growth phases of massive elliptical galaxies \citep{scot,roc,swi06}, and they may 
serve as efficient `lighthouses' of the seeds of massive clusters at $z\gtrsim1.5$ \citep{bla04,far06a,far06b}. 

The still controversial nature of the power source in local ULIRGs, coupled with the central position that ULIRGs seem to play in 
several astrophysical processes at high redshift, makes it ever more important to understand the nature of the heavily obscured 
starburst and AGN activity in these systems. This is ideally done in the mid-infrared, directly sampling the emission from the hot 
dust that shrouds the central engines of ULIRGs, and using fine-structure lines, which suffer much less from extinction effects than 
optical or near-IR lines. The recently launched {\it Spitzer} space telescope \citep{wer04} provides an ideal platform to undertake such studies, with its 
suite of mid-IR instruments, including the Infrared Spectrograph \citep{hou04}, which offers dramatic improvements in 
sensitivity and resolution compared to previous generation facilities. In this paper, we present high resolution mid-infrared spectra of 53 
local ULIRGs, and discuss some spectral diagnostics based on their emission line fluxes and other spectral features. 
We assume a spatially flat cosmology with $H_{0}=70$ km s$^{-1}$ Mpc$^{-1}$, $\Omega=1$, and $\Omega_{m}=0.3$.

\section{Analysis}\label{sectobs}

\subsection{Observations}
The ULIRGs presented here were observed as part of a large study within the IRS GTO program to obtain mid-infrared spectra of 110 low redshift ULIRGs (Spitzer program ID 105). 
These 110 ULIRGs were selected from the IRAS 1Jy \citep{kim98} and 2Jy \citep{str90} spectroscopic surveys, and from the FIRST ULIRG sample \citep{sta00}. 
Low resolution spectra were obtained of all 110 objects, and high resolution spectra were obtained of the 53 brightest (at 60$\mu$m, those with $f_{60}>0.7$Jy) objects. 
The low resolution spectra span 5.2$\mu$m - 38.5$\mu$m, with a resolution of R$\sim60-125$. Initial results are presented in \citet{arm04,spo04} and \citet{arm06}. 
Molecular Hydrogen masses are presented in \citet{hig06}, and crystalline silicate measurements are presented in \citet{spo06}. An atlas of the low resolution 
spectra can be found in \citet{des07}, including Polycyclic Aromatic Hydrocarbon (PAH) luminosities and equivalent widths. Measurements of the strengths of the 
9.7$\mu$m silicate absorption features from the low resolution spectra can be found in \citet{spo07}. 

Here, we present the 53 high resolution spectra. The sample is listed in Table \ref{sample}. A few of the sample have IR luminosities 
that lie slightly below the canonical ULIRG lower limit of 10$^{12}$L$_{\odot}$, but for simplicity we refer to them as ULIRGs for the remainder of this paper. 
Each ULIRG was observed with both the Short-High (SH, 9.9$\mu$m - 19.6$\mu$m, $11.3\arcsec\times4.7\arcsec$, R$\sim600$, 2.3\arcsec\ pix$^{-1}$) and 
Long-High (LH, 18.7$\mu$m - 37.2$\mu$m, $22.3\arcsec\times11.1\arcsec$, R$\sim600$, 4.5\arcsec\ pix$^{-1}$) modules onboard the IRS. The targets were 
placed in the center of each slit by performing 'high' accuracy peak-ups using the blue peak-up array, on either a nearby 2MASS star or on the 
nuclei of the ULIRGs themselves, and observed in two nod positions. For five ULIRGs, their nuclei are separated by distances of $\sim$5\arcsec\ or more; 
for IRAS 08572+3915 the slits were centered on the north-western nucleus, for IRAS 14348-1447 the slits were centered on the south-western nucleus, for 
IRAS 19254-7245 the slits were centered on the southern nucleus, for IRAS 23498+2423 the slits were centered on the north-western nucleus, and for 
Mrk 463 the slits were centered on the eastern nucleus. The other ULIRGs are all either single-nucleus (in ground-based imaging) systems, or have close 
separation ($\lesssim$5$\arcsec$) double nuclei. While the available optical spectroscopy for our sample is not homogenous, there are no examples of 
objects in our sample where two nuclei with clearly different optical spectral classifications fall within the IRS slits. 

For most of the SH observations we observed each object for 6 ramps, 
with a ramp time of 30s, to give a total on-source exposure time of 180s. For the fainter targets we observed for two ramps using a ramp time of 120s, 
for a total on-source exposure time of 240s. The LH observations were the same for each object; namely 4 ramps and a ramp time of 60s, for a total 
on-source exposure time of 240s\footnote{Details for each observation can be found by referencing the AOR keys given in Table \ref{sample} within 
the {\it Leopard} software, available from the Spitzer Science Center}. 

\subsection{Data reduction}
The data were processed through the {\it Spitzer} Science Center's pipeline reduction software (version 13.2), which performs standard reduction 
tasks such as ramp fitting and dark current subtraction.  To ensure an accurate flatfielding correction we started our reduction from the {\it un}flatfielded ({\it droopres})
images. Starting with these frames, we flagged rogue and otherwise `bad' pixels using the {\em irsclean}\footnote{This tool is available from the SSC
website: http://ssc.spitzer.caltech.edu} tool, which uses a mask of rogue pixels for each campaign to first flag and then replace rogue
pixels. The individual frames were then combined into a single image, and spectra were extracted from each nod using the SMART software package
\citep{hig}, using full-slit extraction. Wavelength and flux calibration were performed by dividing the extracted spectra by that of a standard 
star, $\xi$\,Dra, and multiplying by its template \citep{coh03}. Features present in only one nod position were treated as
artifacts and removed manually. The two nods were then combined. The pixels on the edge of each order (typically the first and last 12 pixels) 
corresponding to regions of decreased sensitivity on the array were then removed to give the final spectrum for each object. The resulting spectra were generally 
of excellent quality. In a few cases some slight mismatch in continuum fluxes between orders was apparent, but not to an extent that could impact 
the analysis of emission line fluxes. 

Both the SH and LH slits are too small to allow for on-slit background subtraction. Ideally the sky continuum background (which is comprised mainly of 
temporally and spatially varying zodiacal light) should be subtracted using contemporaneous `sky' observations, taken with similar exposure times 
and as close on the sky as possible to the observations of the target. Such observations were however not taken for our sample. In the absence of sky 
observations, modelled sky fluxes can be used \citep{rea03}, but they are uncertain by at least a factor of two. In this paper we are 
interested only in the emission line fluxes, for which background continuum subtraction is not necessary. Hence, we do not correct our spectra for 
contamination from sky continuum background. 

We measured line fluxes and wavelengths by assuming the emitting region was a point source at the spatial resolution of the SH and LH slits, subtracting the continuum via a cubic spline fit over a $\sim0.5\mu$m region centered on each line, and then fitting a single Gaussian profile to each line. We found a pure Gaussian profile to be a good fit to the lines in virtually all cases. The resulting 3$\sigma$ 
uncertainties in the wavelengths are themselves a function of wavelength, and are typically 0.01$\mu$m at 10$\mu$m, 0.03$\mu$m at 20$\mu$m, and 
0.04$\mu$m at 30$\mu$m. Blended lines were measured by simultaneously fitting multiple gaussians to the combined profile. Upper limits were 
determined by measuring the noise level of the data at the wavelength where the line is expected to lie; this can give rise to significant 
variations in upper limits for different lines in the same source if the wavelength of a line lies in an order overlap region.

\subsection{Extinction corrections}\label{noextinct}
Previous authors have generally used either near-IR line ratios, or the [SIII]$\lambda\lambda$18.713,33.481 line ratio to derive 
an extinction correction (e.g. \citealt{ver}), however we only detect both [SIII] lines in a small fraction of our sample, and the available near-IR spectroscopy 
for our sample is heterogenous and incomplete. Given this, and the uncertain structure of the narrow-line emitting gas in the mid-IR, we have chosen not 
to correct our line fluxes for extinction. We can however estimate what effect this lack of an extinction correction will have on our analysis. The relationship 
between the intrinsic and observed flux ratio for a pair of emission lines at wavelengths $\lambda_{1}$ and $\lambda_{2}$ can be written:

\begin{equation}\label{extincteqn}
\left(  \frac{I_{\lambda_{1}}}{I_{\lambda_{2}}}\right)_{int} = \left(  \frac{I_{\lambda_{1}}}{I_{\lambda_{2}}}\right)_{obs}\times10^{ 0.4\left( \frac{A_{\lambda_{1}}}{A_{V}} - \frac{A_{\lambda_{2}}}{A_{V}} \right)A_{V} }
\end{equation}

\noindent where $A_{\lambda_{1}}$, $A_{\lambda_{2}}$ and $A_{V}$ are the extinctions at the wavelengths of the pair of lines and in the rest-frame $V$ band, 
respectively. Assuming $A_{V} = 3.169 \times E(B-V)$, and obtaining values for $A_{\lambda}/E(B-V)$ from a standard extinction law \citep{li01} then allows 
us to estimate the effect on a line ratio in terms of an increase in the $V$ band extinction. Using this formalism, we 
find that the effect on most line ratios if $A_{V}$ is increased is small. For example, for the [NeIII]/[NeII] ratio we obtain:

\begin{equation}
\left(  \frac{[NeIII]}{[NeII]}\right)_{int} = \left(  \frac{[NeIII]}{[NeII]}\right)_{obs}\times10^{\left( -7.34\times10^{-4} \right)A_{V} }
\end{equation}

\noindent which is negligible unless the increase in $A_{V}$ is at least a few tens. Hence our lack of an extinction correction, while important 
to be aware of, should not unduly affect our analysis of most line ratios, though the effect of a lack of an extinction correction on line 
luminosities is more significant. We discuss the effects of extinction in terms of an increase in $A_{V}$ for both fine structure line ratios 
and luminosities in the following sections.

Finally, we note that the magnitude, and in some cases the direction, of the effect on a line ratio for an increase in extinction depend significantly on 
ones choice of extinction law. In Table \ref{extinctlaws} we present the scaling factors for several line ratios for an increase in $A_{V}$, assuming some
commonly used extinction laws. For a given line ratio there is broad consistency between the extinction laws, but differences of up to $25\%$ between the 
scaling factors are common. These differences in scaling factors between extinction laws should be kept 
in mind in the following sections.

\section{Results}
\begin{figure}
\begin{minipage}{170mm}
\includegraphics[angle=90,width=150mm]{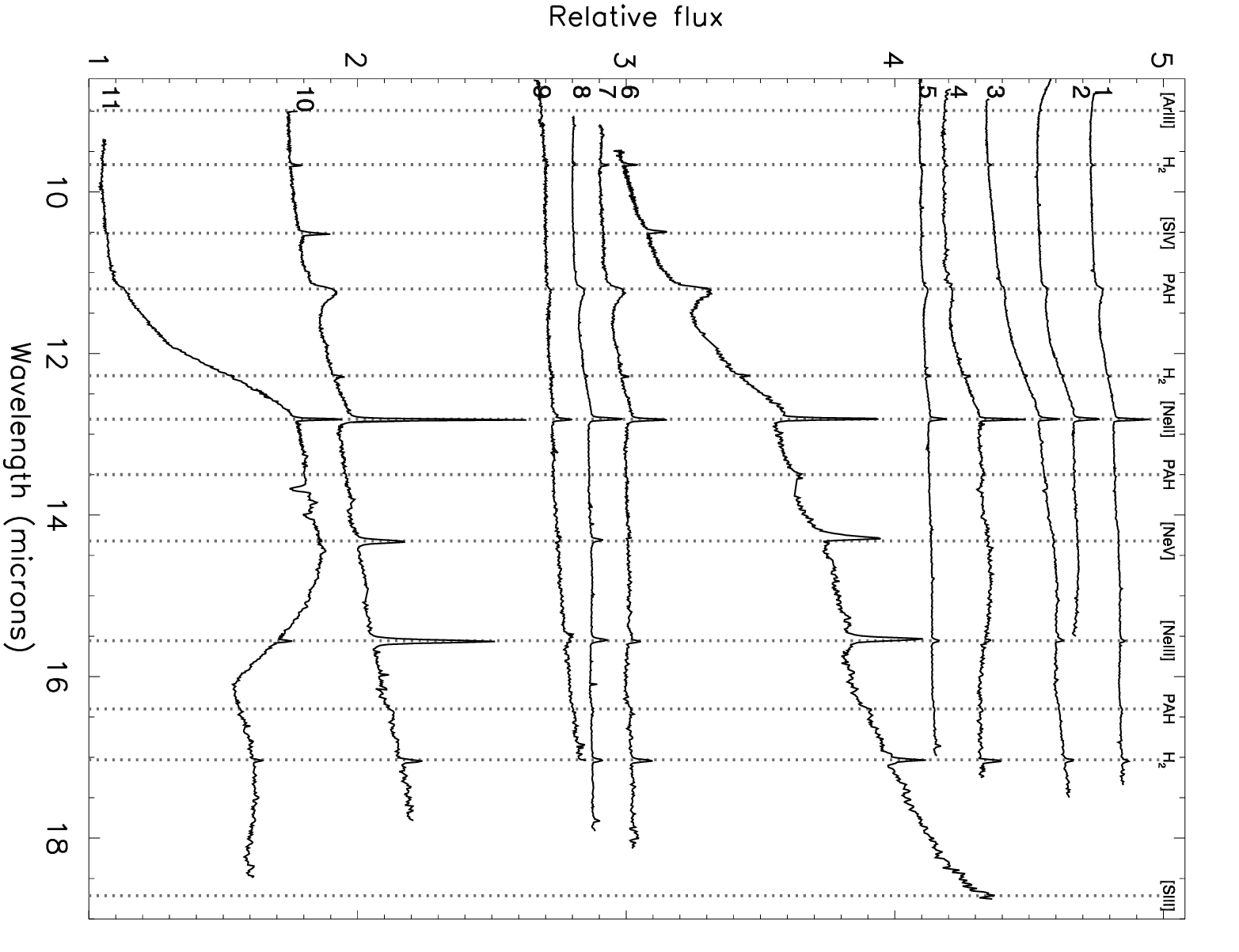}
\end{minipage}
\caption{Short-High spectra of our sample. The numbers on the left hand side correspond to the ID numbers in column one of Table \ref{sample}. Wavelengths have been 
shifted to the (optical) rest-frame. Some objects also show absorption at 13.7$\mu$m and 14.0$\mu$m, caused by C$_{2}$H$_{2}$ and HCN gas \citep{lah07}. \label{sh_figa}}
\end{figure}

\begin{figure}
\begin{minipage}{170mm}
\includegraphics[angle=90,width=150mm]{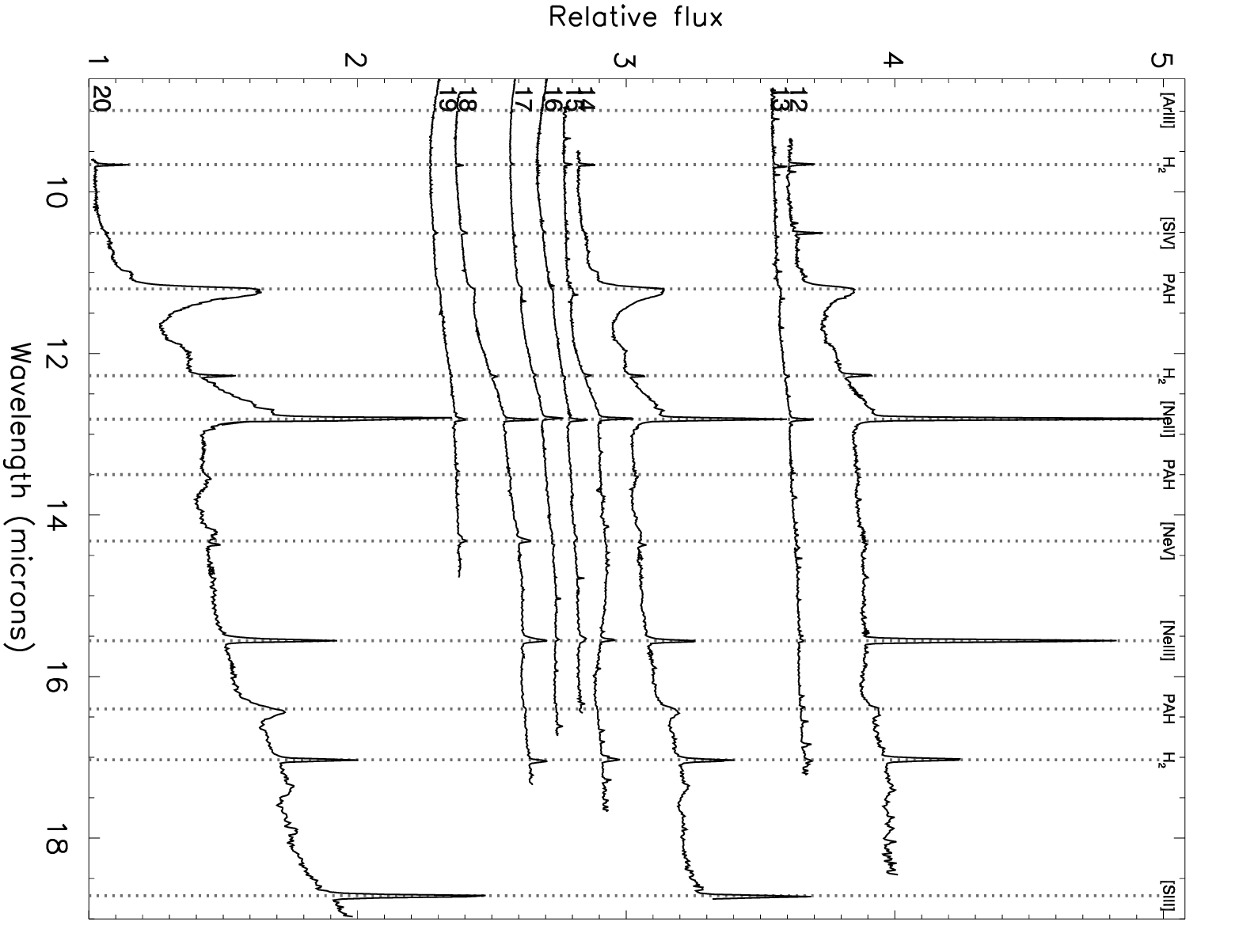}
\end{minipage}
\caption{Short-High spectra (continued) \label{sh_figb}}
\end{figure}

\begin{figure}
\begin{minipage}{170mm}
\includegraphics[angle=90,width=150mm]{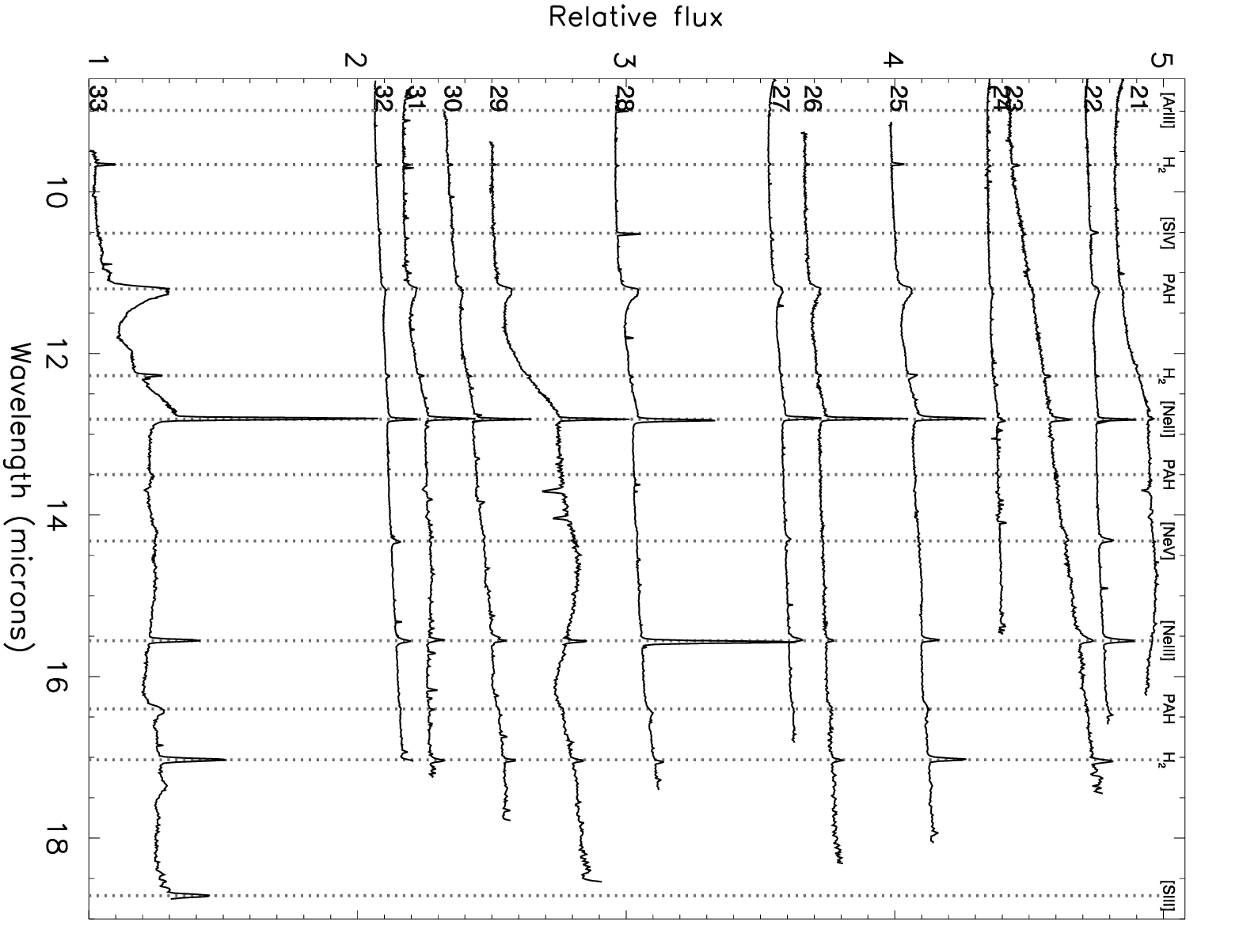}
\end{minipage}
\caption{Short-High spectra (continued) \label{sh_figc}}
\end{figure}

\begin{figure}
\begin{minipage}{170mm}
\includegraphics[angle=90,width=150mm]{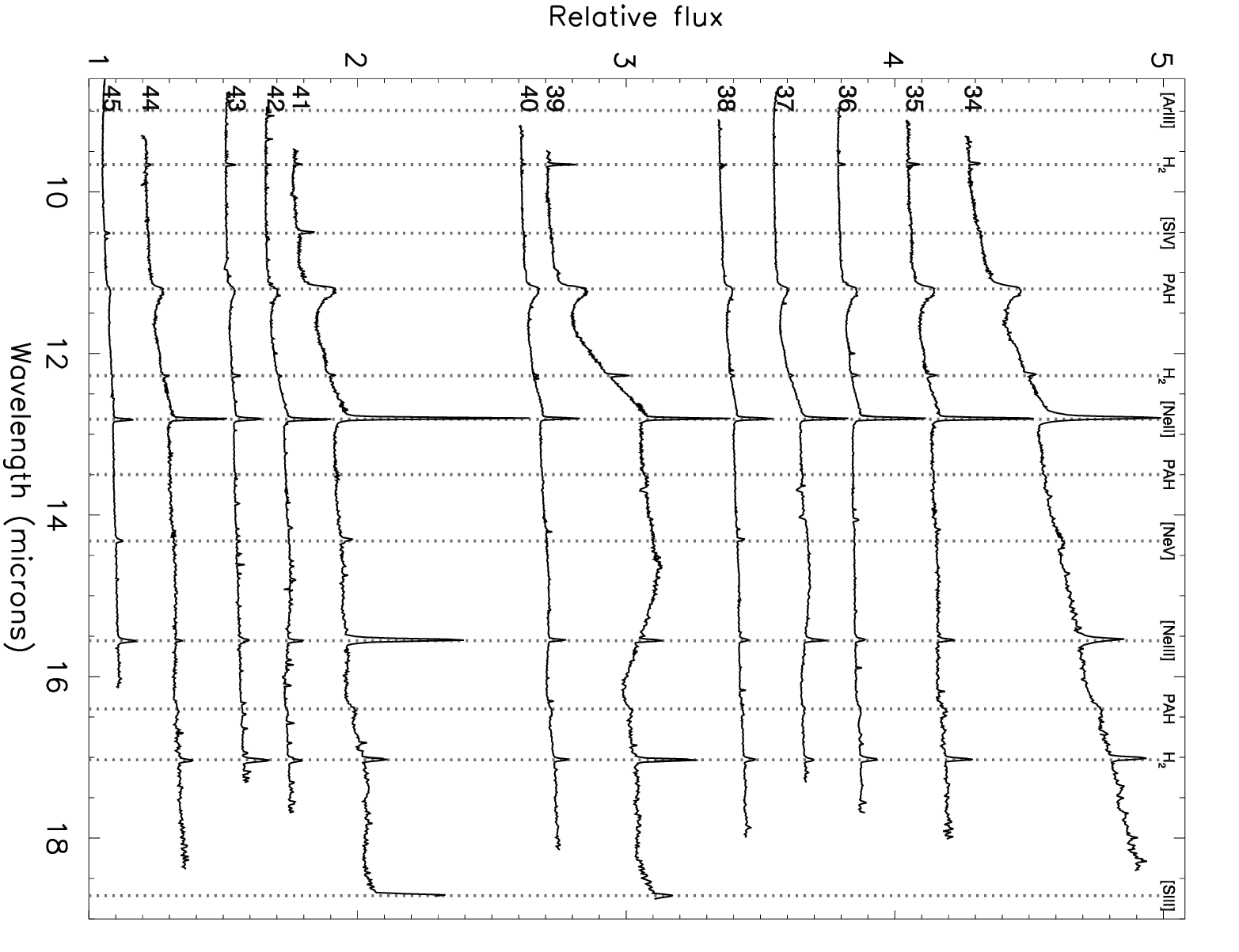}
\end{minipage}
\caption{Short-High spectra (continued) \label{sh_figd}}
\end{figure}

\begin{figure}
\begin{minipage}{170mm}
\includegraphics[angle=90,width=150mm]{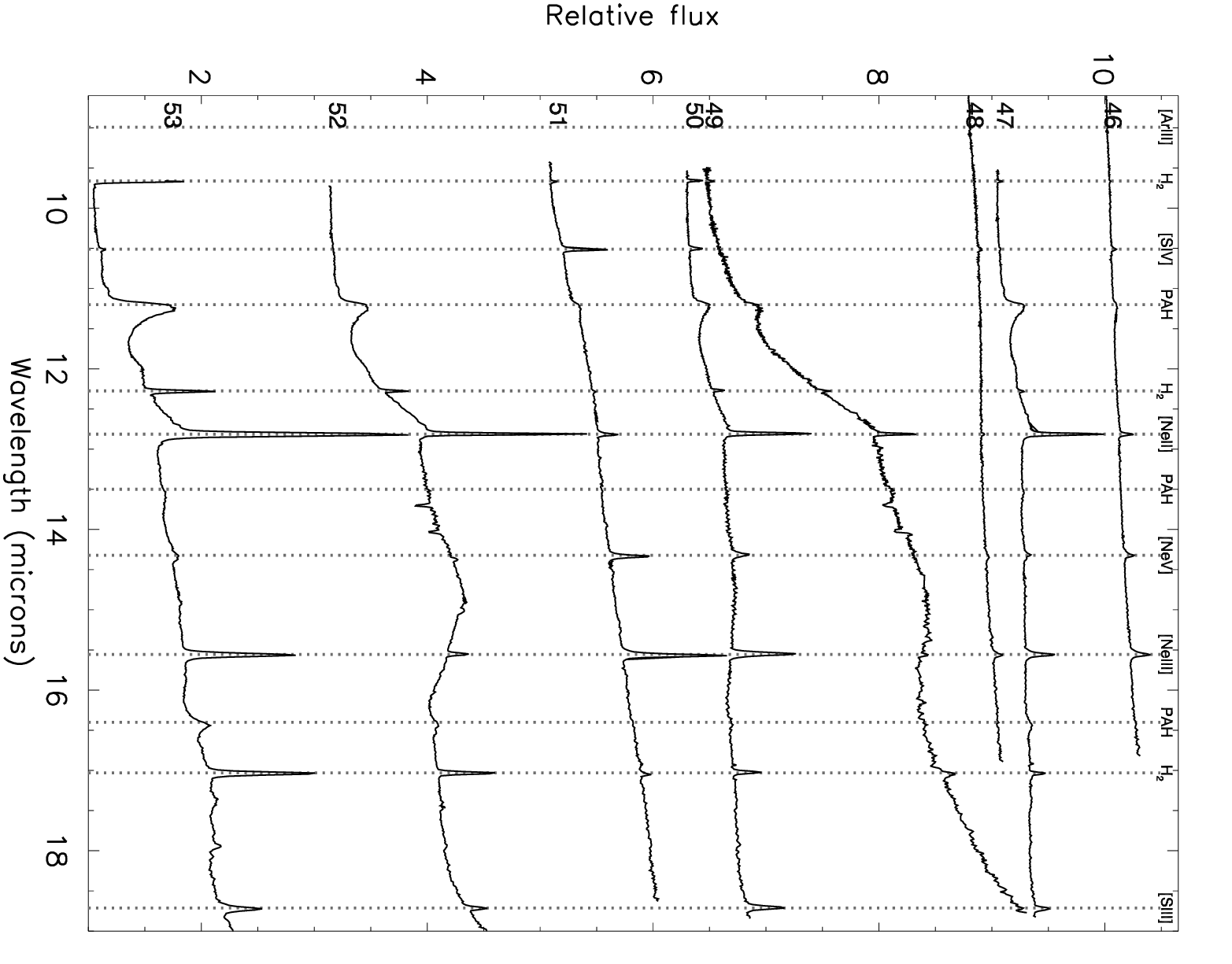}
\end{minipage}
\caption{Short-High spectra (continued). Note different $y$ axis scaling.  \label{sh_fige}}
\end{figure}

\begin{figure}
\begin{minipage}{170mm}
\includegraphics[angle=90,width=150mm]{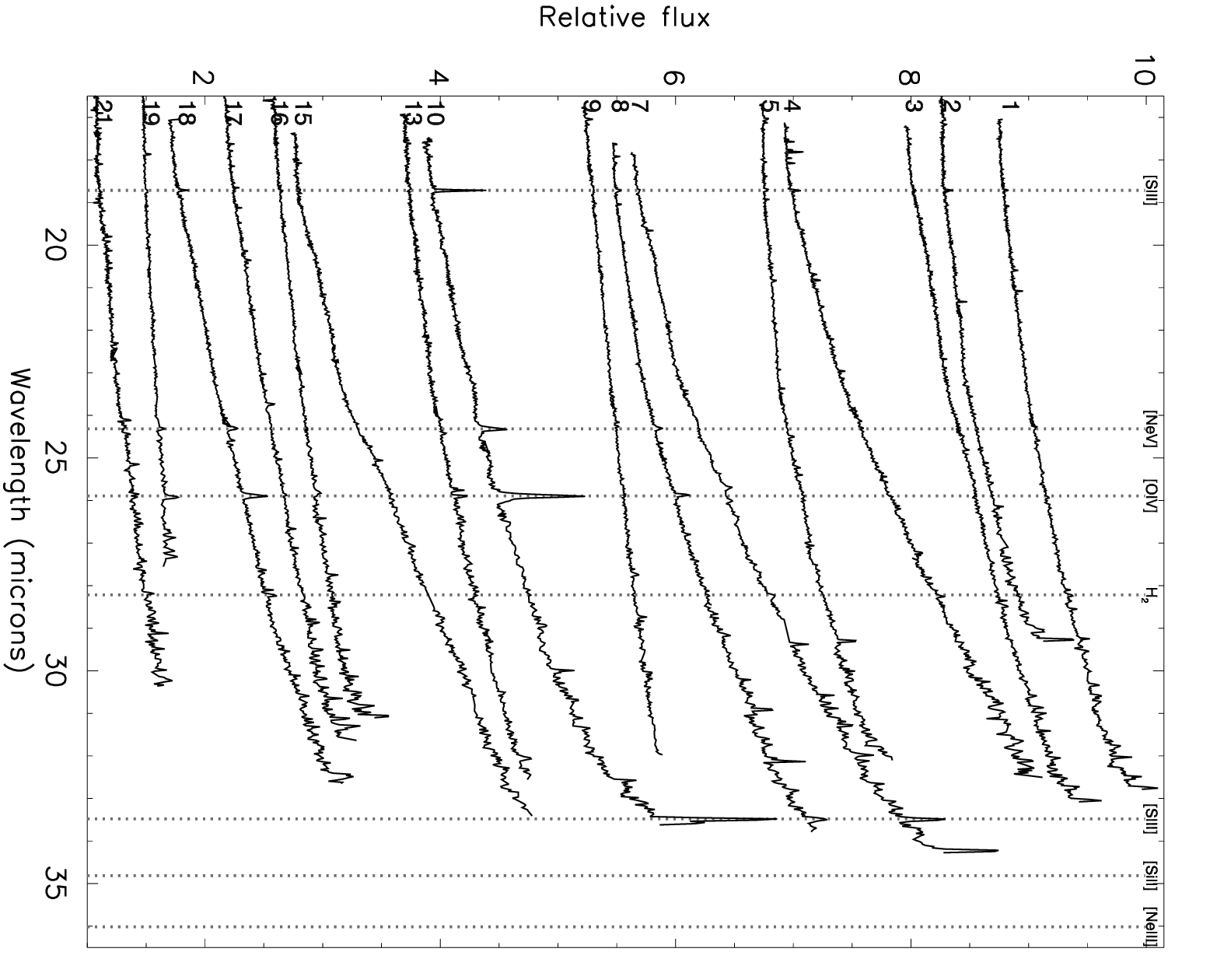}
\end{minipage}
\caption{Long-High spectra of our sample. 
Note that the LH spectra are not arranged in order of ID number from Table \ref{linefluxa} due to the wide dispersion in their continuum slopes.  
Wavelengths have been shifted to the (optical) rest-frame. Some objects also show an OH absorption feature at rest-frame 34.6$\mu$m. \label{lh_figa}}
\end{figure}

\begin{figure}
\begin{minipage}{170mm}
\includegraphics[angle=90,width=150mm]{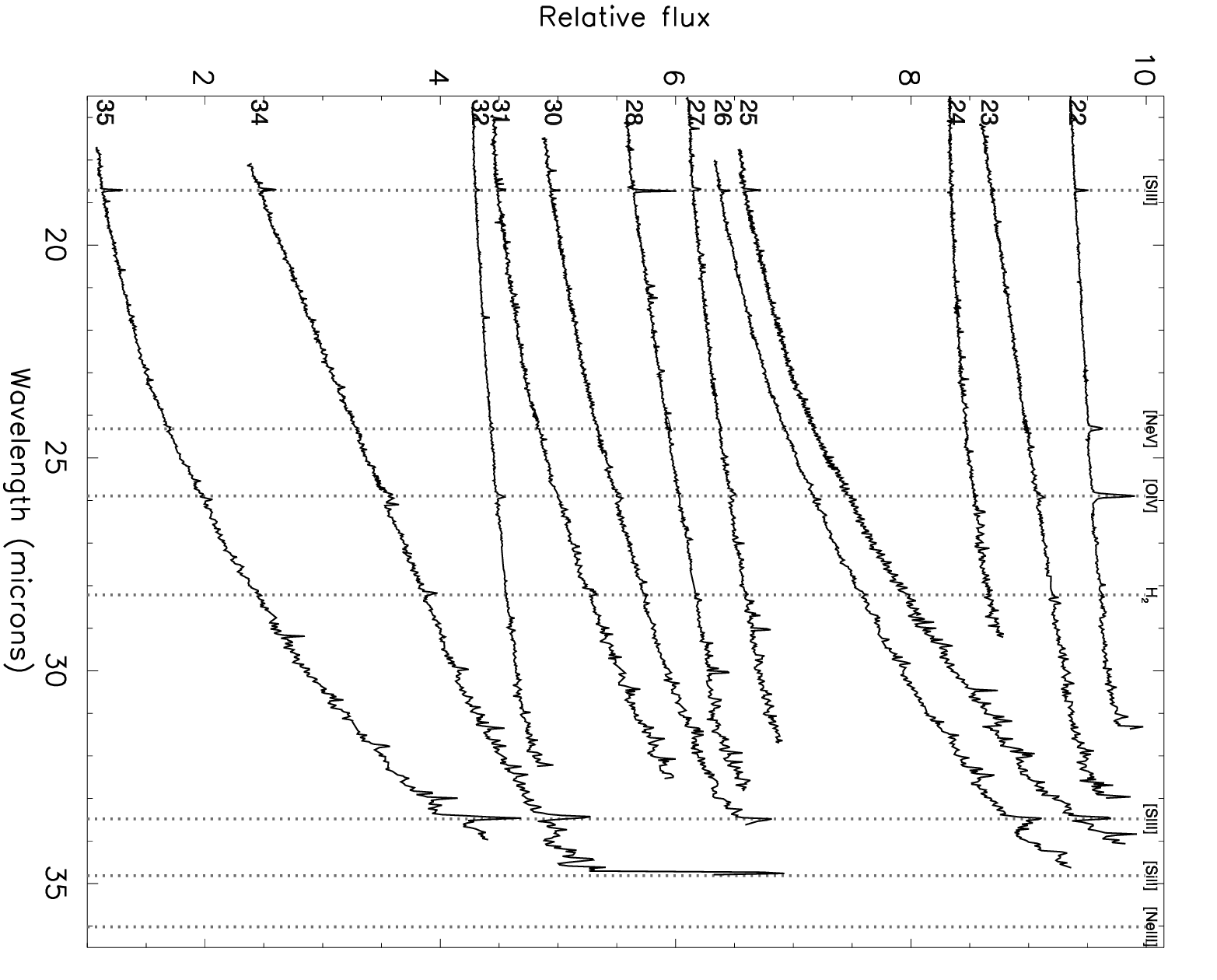}
\end{minipage}
\caption{Long-High spectra (continued) \label{lh_figb}}
\end{figure}

\begin{figure}
\begin{minipage}{170mm}
\includegraphics[angle=90,width=150mm]{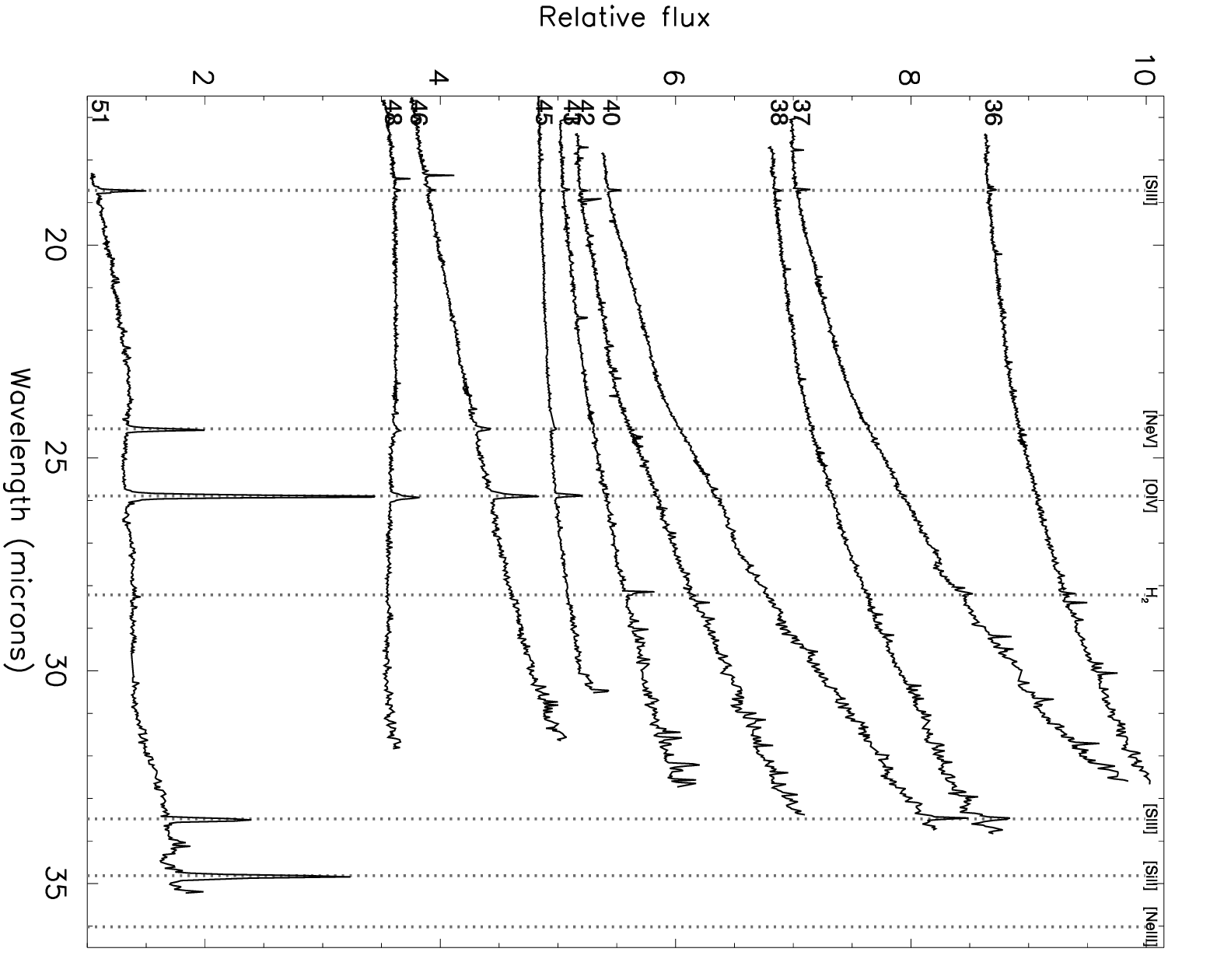}
\end{minipage}
\caption{Long-High spectra (continued)  \label{lh_figc}}
\end{figure}

\begin{figure}
\begin{minipage}{170mm}
\includegraphics[angle=90,width=150mm]{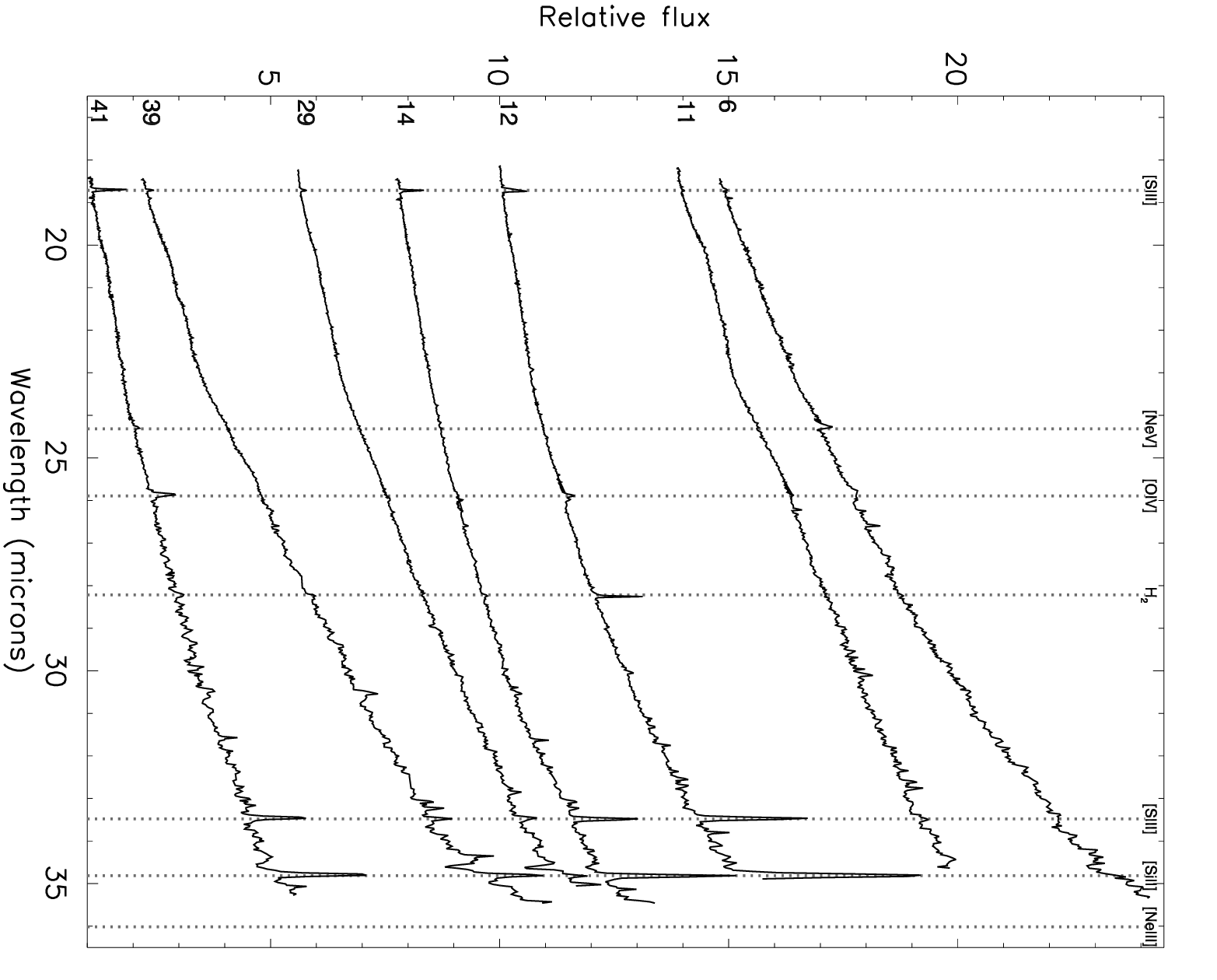}
\end{minipage}
\caption{Long-High spectra (continued). Note that the y axis scaling in this plot differs from that in Figures \ref{lh_figa}, \ref{lh_figb} and \ref{lh_figc}. \label{lh_figd}}
\end{figure}

\begin{figure}
\begin{minipage}{170mm}
\includegraphics[angle=90,width=150mm]{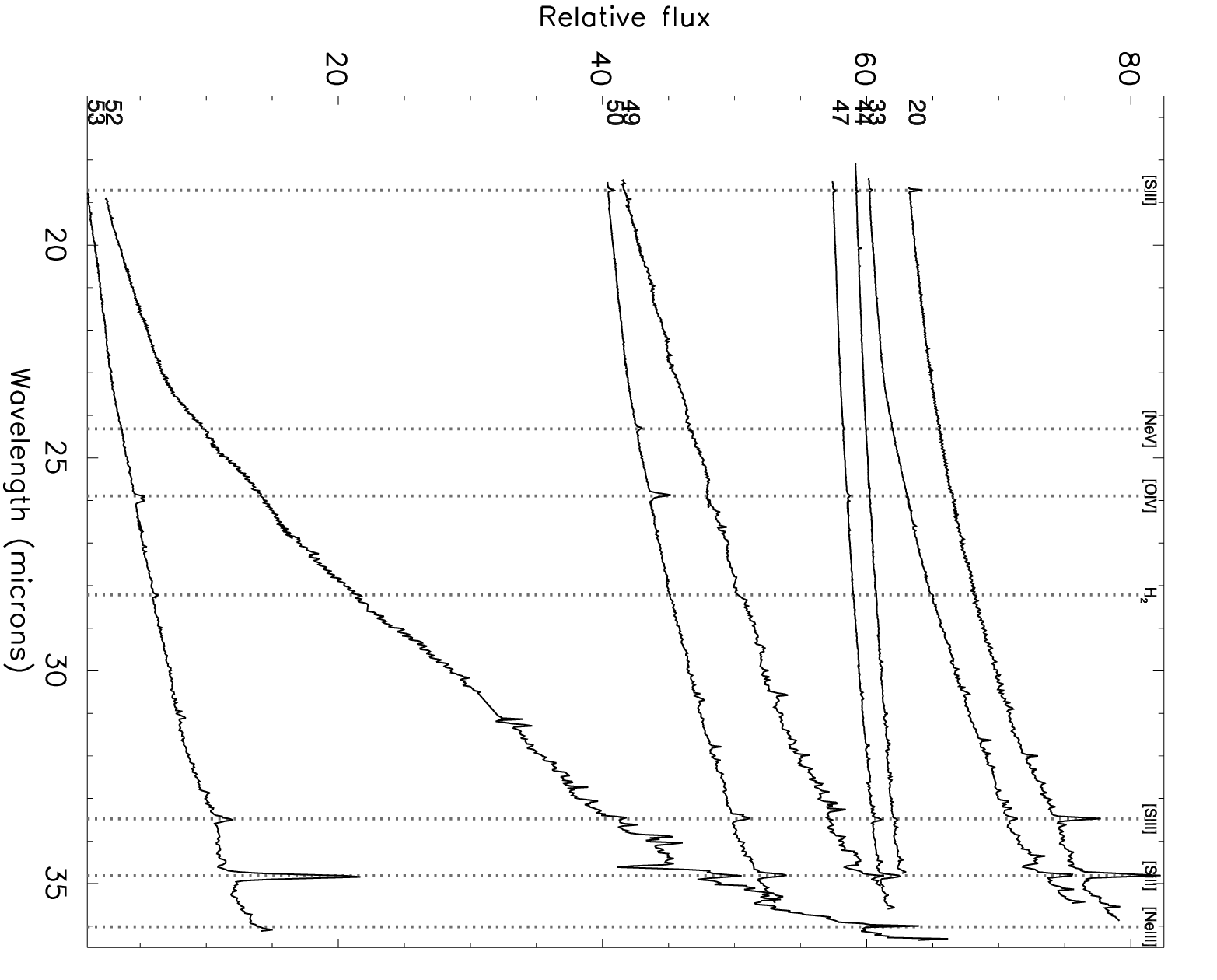}
\end{minipage}
\caption{Long-High spectra (continued). Note that the y axis scaling in this plot differs from that in Figures \ref{lh_figa}, \ref{lh_figb}, \ref{lh_figc} and \ref{lh_figd}.  \label{lh_fige}}
\end{figure}

The SH spectra are presented in Figures \ref{sh_figa} to \ref{sh_fige}, and the LH spectra are presented in Figures \ref{lh_figa} to \ref{lh_fige}. 

\subsection{Common lines}
All of the spectra show various fine structure emission lines of Neon, Oxygen, Sulfur, Silicon and (depending on redshift) Argon. Also present are 
molecular Hydrogen lines. These lines are listed in Table \ref{linefluxa}. We detect [NeII]$\lambda$12.81 and [NeIII]$\lambda$15.56 in nearly 
every object. [SIII]$\lambda$18.71 is also common, detected in $\sim$80\% of the sample, and [SIV]$\lambda$10.51 is detected in just under 
half of the sample. The detection of [SIII]$\lambda$33.48 is dependent on redshift, requiring $z\lesssim 0.11$ (for the line to lie at $<37\mu$m in 
the observed frame), but is detected in $\sim$80\% of the objects where this line lies within the LH bandpass. Other lines 
whose detection is dependent on redshift include [ArII]$\lambda$8.99, which is detected in the majority of objects at $z\gtrsim 0.14$, and [SiII]$\lambda$34.82, 
which is seen in about half of the objects at $z\lesssim 0.06$ (though this line lies in a noisy LH order even at $z=0$ and thus is hard 
to detect). Three higher ionization lines are also present, though less commonly than the lines discussed above; [NeV]$\lambda$14.32 
and/or [OIV]$\lambda$25.89 are detected in just under half the sample, while [NeV]$\lambda$24.32 is detected in about one third of the sample. 
Turning to molecular Hydrogen lines; the S(3), S(2) and S(1) pure rotational transitions of H$_{2}$ are seen in nearly all of the sample, while 
the S(0) H$_{2}$ transition is seen in about one third of the sample, though its rarity compared to the other molecular Hydrogen lines is probably 
as much to do with the rising continuum towards longer wavelengths as anything else. Our line fluxes are in all cases consistent with those in 
\citet{arm04,arm06} and \citet{hig06}, though we use a more recent version of the IRS pipeline. The fluxes reported here should therefore be more 
   accurate than those previously published.

In addition to fine-structure (and other) emission features, some objects show one or more absorption features. The focus of this paper is the 
fine-structure emission lines, so here we only briefly mention these features, deferring a full discussion to later work. Several objects show two absorption features, 
corresponding to the vibration-rotation absorption bands of $C_{2}H_{2}$ at 13.70$\mu$m and HCN at 14.02$\mu$m. An extensive discussion of these features can be found 
in \citet{lah07}. A total of ten objects, most prominently Arp 220, IRAS 15250+3609 and IRAS 20551-4250, show an OH absorption feature at $\sim$34.6$\mu$m 
(IRAS 15250+3609 may also show a further OH absorption feature at rest-frame 28.9$\mu$m, though the significance of detection is weak). This feature is most 
likely the $^{2}\Pi_{3/2}J=3/2 - ^{2}\Pi_{1/2}J=5/2$ OH absorption doublet, which is thought to pump the 1667MHz OH maser line, among others \citep{eli76}. This absorption 
feature has been seen previously in Galactic sources (e.g. \citealt{jus96}), NGC 253 \citep{goi05}, and Arp 220 \citep{ski97}. The prominence of this feature in Arp 
220 is consistent with the presence of an OH megamaser in this source \citep{lon94}.

\subsection{Unusual lines} \label{rarelines}
In addition to the lines discussed in the previous section, we also see a variety of `rare' (which we arbitrarily define as appearing in ten or fewer objects) 
emission features in several objects, listed in Table \ref{linefluxc}. We detect [ClII]$\lambda$14.37 in six objects, four of which also show [NeV]$\lambda$14.32, leading 
to a double-peaked profile (see also \citealt{arm06a}). In the other two objects only one peak is seen, and it is possible that we have confused [ClII]$\lambda$14.37 with 
[NeV]$\lambda$14.32, however the velocity shift relative to the systemic (optical) redshift would have to be $>5000$km s$^{-1}$ for this to be the case. We therefore regard 
[ClII]$\lambda$14.37 as the more likely identification. This line is seen in some lower luminosity starbursts \citep{spo00,ver}, but is rare in AGN \citep{stu02}. A number 
of low-ionization iron lines are present, including [FeII]$\lambda\lambda\lambda$17.94,24.52,25.99 and [FeIII]$\lambda$22.93, all of which are seen (rarely) in lower 
luminosity starbursts and AGN. We tentatively detect [PIII]$\lambda$17.89 in two objects. This line is seen in small numbers of local IR-luminous sources, and 
in a variety of Galactic sources. Also present is the [ArV]$\lambda$13.10 line, which is occasionally seen in AGN \citep{stu02}. One object, Arp220, shows a weak detection of what is plausibly [NeIII]$\lambda$36.01, though Arp220 is the only object in our sample at a low enough redshift for 
this line to enter a well behaved part of the LH bandpass. We weakly detect the HI 7-6 line (i.e. 
the alpha transition of the Humphreys atomic Hydrogen series) in four objects. Finally, two objects show features that appear to be real, but proved difficult to identify 
reliably, hence we have not listed them in Table \ref{linefluxc}. IRAS 23498+2423 shows a feature at a rest-frame wavelength of 10.581$\mu$m with a flux of 
3.17$\times 10^{-22}$ W cm$^{-2}$. This feature is clearly resolved from the [SIV]$\lambda$10.511 line. If this feature is real, then possibilities include [CoII]$\lambda$10.521, 
which is seen in the 200 day - 500 day mid-IR spectra of core collapse supernovae \citep{kot}, or [NiII]$\lambda$10.682. IRAS 19297-0406 shows an emission feature with a 
flux of 3.88$\times10^{-21}$ W cm$^{-2}$ that can plausibly be identified as the 31.77$\mu$m water ice emission feature.

\section{Discussion}

\subsection{Properties of the narrow-line region gas}
Mid-IR fine structure lines can be used to study three properties of the narrow-line region (NLR) gas\footnote{Due to Spitzers limited spatial resolution, the `NLR' of a 
ULIRG should be regarded as the ensemble of all the regions in a ULIRG where gas heated by star formation and/or an AGN emits fine-structure lines in the mid-IR}; excitation, 
electron temperature and electron density. Measuring electron temperature though requires combining a mid-IR line with an optical or near-IR line of the same species. In 
this paper, we therefore concentrate only on electron density and excitation.

\subsubsection{Electron Density} \label{elecdens}
The electron density in a NLR can be estimated by using 
the ratios of two lines representing transitions from the same orbital and suborbital (i.e. the same principal and angular momentum quantum numbers), but with different 
numbers of electrons in that (sub)orbital. For our sample there are two line ratios that satisfy this requirement; the $^{3}p_{2} - ^{3}p_{1}$ and $^{3}p_{1} - ^{3}p_{0}$ 
transitions of [NeV] at 14.32$\mu$m and 24.32$\mu$m, and the $^{3}p_{2} - ^{3}p_{1}$ and $^{3}p_{1} - ^{3}p_{0}$ transitions of [SIII] at 18.71$\mu$m and 33.48$\mu$m. These ratios 
can then be converted to an electron density by solving the relevant rate equation. Of the two, the [NeV] ratio is more useful. Its high ionization potential of 97eV means 
that it can only be produced (at a level observable in a ULIRG) by an AGN, making it more straightforward to interpret than the [SIII] ratio. Furthermore, for most extinction 
laws the extinctions of the two [NeV] lines are virtualy identical (Table \ref{extinctlaws}, see also \citealt{dra03}), making the [NeV] ratio only marginally sensitive to the lack of an extinction correction. 

Sixteen of the ULIRGs in our sample show detections in both [NeV] lines. The line ratios for these objects span the range 0.57 $<$ [NeV]$\lambda$14.32/[NeV]$\lambda$24.32 $<$ 2.69, 
with a fairly even spread. The nine objects that are only detected in [NeV]$\lambda$14.32 have lower limits on the [NeV] line ratio consistent with this range. Using Figure 3 of \citet{ale99}, then these [NeV] line ratios are consistent with electron densities of $<10^{4}$ cm$^{-3}$ in all cases, well below the critical densities of the two lines. Our derived electron densities are also comparable to those derived (using the same [NeV] ratio) for lower luminosity AGN \citep{stu02}. As a check, we examine the [SIII] line ratio; for those objects with detections in both lines we see a range of 0.14 $<$ [SIII]$\lambda$18.71/[SIII]$\lambda$33.48 $<$ 1.10, consistent with electron densities of $\lesssim10^{3.5}$cm$^{-3}$. 
It is important to note however that we expect the [NeV] and [SIII] line ratios to give different electron densities, because the [NeV] lines are likely produced 
solely by the AGN, while the [SIII] lines (probably) arise from both starburst and AGN heated regions. 

\subsubsection{Excitation} \label{elecexcite}
If electron densities are below the critical density, then the excitation level of the NLR can be estimated by considering flux ratios of adjacent ionization 
states of the same element, e.g. for an element $X$; $f_{X^{i+1}}/f_{X^{i}}$. For a fixed number of ionizing photons per Hydrogen atom (i.e. a given ionization parameter $U$), this 
ratio will be approximately proportional to the number of photons producing the observed $X^{i}$ flux relative to the number of Lyman continuum photons, though if $U$ varies then $f_{X^{i+1}}/f_{X^{i}}$ will also vary. Therefore, for a fixed $U$, a higher value of $f_{X^{i+1}}/f_{X^{i}}$ indicates a harder radiation field, though the details depend on the element in question. 

Our spectra contain a limited number of lines, hence detailed excitation diagnostics are not possible. We can however get a qualitative idea of the range in 
excitations in our sample. The two most useful diagnostic ratios are [NeIII]$\lambda$15.56/[NeII]$\lambda$12.81 and [SIV]$\lambda$10.51/[SIII]$\lambda$18.71. 
Both Neon and Sulfur are abundant in Galactic sources, with both [Ne/H] and [S/H] lying approximately in the range $10^{-4} - 10^{-6}$. Furthermore, the photon 
energies required to produce any of these four ions are $<50$eV, meaning that they can all be produced in star-forming regions as well as AGN. All four lines are 
seen in Wolf-Rayet star spectra  \citep{smi01}, HII regions \citep{pee02}, planetary nebulae \citep{ber01},  as well as the integrated spectra of 
local galaxies. We might also expect [NeIII] and [SIV] to increase in strength relative to [NeII] and [SIII] in lower density and/or lower metallicity star 
forming regions \citep{tho06,ho07}. It is however strange, given their similar ionization energies, that we detect [NeIII]$\lambda$15.56 
in virtually all the sample, but only detect [SIV]$\lambda$10.51 in about 45\% of the sample. The [SIV]$\lambda$10.51 line lies close to the 9.7$\mu$m 
silicate feature, but we detect the H$_{2}$,S(3)$\lambda$9.66 line in most of our sample, in many cases with a flux that is lower than the upper limit on the 
[SIV]$\lambda$10.51 line. The most likely explanation for this is that the Neon and Sulfur emitting zones in ULIRGs lie within regions extincted by 
silicate dust, while the H$_{2}$ emitting region lies outside it (see also \citealt{hig06}). This is supported by the fact that we do not detect 
[SIV]$\lambda$10.51 in any object which has a silicate strength \citep{lev07} greater than 2.1 (see \citet{spo07} for the silicate strengths of our sample).

\begin{figure}
\begin{minipage}{170mm}
\includegraphics[angle=90,width=150mm]{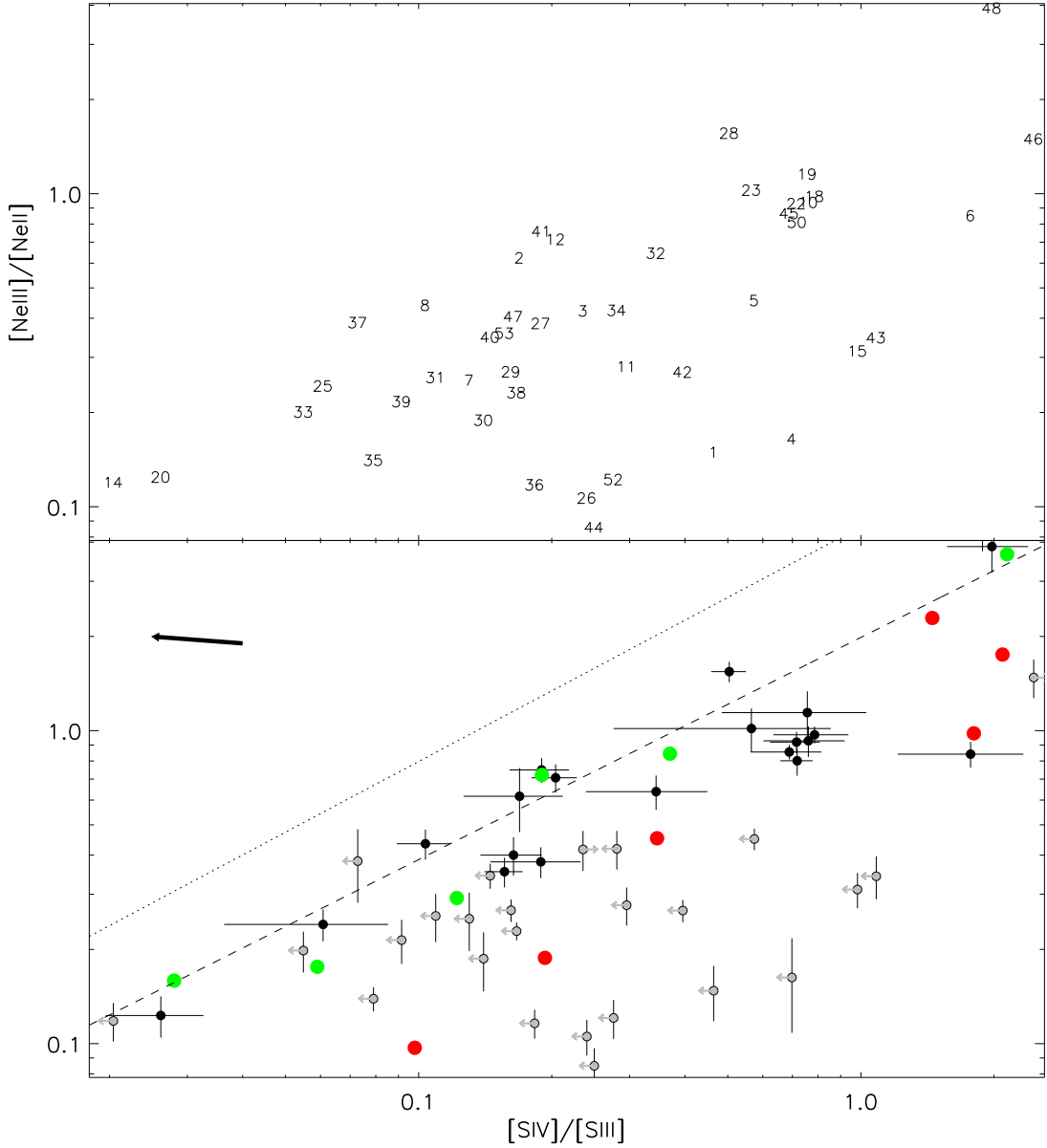}
\end{minipage}
\caption{Excitation diagnostic. The top panel identifies the ULIRGs using the numbers in column 1 of Table \ref{sample}, while 
the bottom panel shows the ULIRGs with error bars and $3\sigma$ limits, as well as ancilliary data. ULIRGs with detections on both 
axes are plotted in black, while ULIRGs with limits on one or both axes are plotted in grey. Green symbols are starbursts \citep{ver} 
and red symbols are AGN \citep{stu02}, both with IR luminosities approximately in the range $10^{10} - 10^{11.5}$L$_{\odot}$. The 
dotted line is the fit to star-forming galaxies, while the dashed line is a fit to Seyfert galaxies, both with IR luminosities 
of $\lesssim10^{10}$L$_{\odot}$ \citep{dal06}. The arrow indicates the effect on a points position if the V band extinction is 
increased by $A_{V}=30$ (but see Table \ref{extinctlaws}).  \label{excite}}
\end{figure}

In Figure \ref{excite} we plot an `excitation plane' of [NeIII]/[NeII] vs [SIV]/[SIII] for our sample. The ULIRGs span a broad range in excitation, but 
with a positive correlation between the Sulfur and Neon line ratios, suggesting that we are seeing emission from the same region in both species. A similar 
correlation has previously been noted for Blue Compact Dwarf galaxies \citep{wu06}, and for nearby infrared-faint galaxies \citep{dal06}. The ULIRGs are fairly 
evenly distributed across the correlation region, with no obvious over- or underpopulated regions. 

Also plotted in Figure \ref{excite} are data for starbursts and AGN with IR luminosities (mostly) between $1\times10^{10}$L$_{\odot}$ and $3\times10^{11}$L$_{\odot}$ 
\citep{ver,stu02}, and  the relations from figure 4 of \citet{dal06} for star forming regions and IR-faint AGN, the bulk of which have IR luminosities below 
$1\times10^{10}$L$_{\odot}$. The ULIRGs occupy the same region in the ionization plane as the \citet{ver} and \citet{stu02} starburst and AGN samples. Barring 
some particularly subtle bias from a lack of extinction correction, this suggests that the mid-IR emitting narrow line region gas in ULIRGs is excited in broadly 
the same way as the NLR in $10^{10}\lesssim$ L$_{ir}$(L$_{\odot}$) $\lesssim3\times10^{11}$ starbursts and AGN, and that any differences in density, metallicity or star formation history between ULIRGs and lower luminosity systems are not sufficient to manifest themselves in simple ionization plane diagrams. 

The same does not however appear to be true for systems with L$_{ir}\lesssim10^{10}$L$_{\odot}$. The ULIRGs are consistent with the slopes of the \citet{dal06} relations, but are offset below them by $\sim$0.2 dex. Increased extinction in the form of a foreground screen could explain this, but would require an additional $A_{V}\simeq60$ of foreground extinction (and see also Table \ref{extinctlaws}). An (arguably) more likely explanation is that there is a decrease in the intrinsic [NeIII]/[NeII] ratio for a given [SIV]/[SIII] ratio in going from L$_{ir}\simeq10^{10}$L$_{\odot}$ starbursts to L$_{ir}\simeq10^{12}$L$_{\odot}$ systems. It is beyond the scope of this paper to investigate this effect in detail, so here we simply suggest a possible explanation. From Table \ref{linefluxa}, [SIV] has a significantly smaller ionization energy than [NeIII], and the difference in ionization energies between [SIV] and [SIII] is smaller than the difference between [NeIII] and [NeII]. This means that, if we increase the gas density in the NLR or decrease the hardness of the ionizing radiation then both the [NeIII]/[NeII] and [SIV]/[SIII] ratios will decrease, but the [NeIII]/[NeII] ratio will decrease by a larger fraction than the [SIV]/[SIII] ratio. As we might expect the gas density in the NLR of ULIRGs to be significantly higher than in L$_{ir}\lesssim10^{10}$L$_{\odot}$ star forming regions, we might also expect a smaller [NeIII]/[NeII] ratio in ULIRGs for a given [SIV]/[SIII] ratio than in lower luminosity starbursts. A softer ionizing radiation field in ULIRGs compared to lower luminosity starbursts would produce a similar effect. 

\begin{figure}
\begin{minipage}{160mm}
\includegraphics[angle=90,width=150mm]{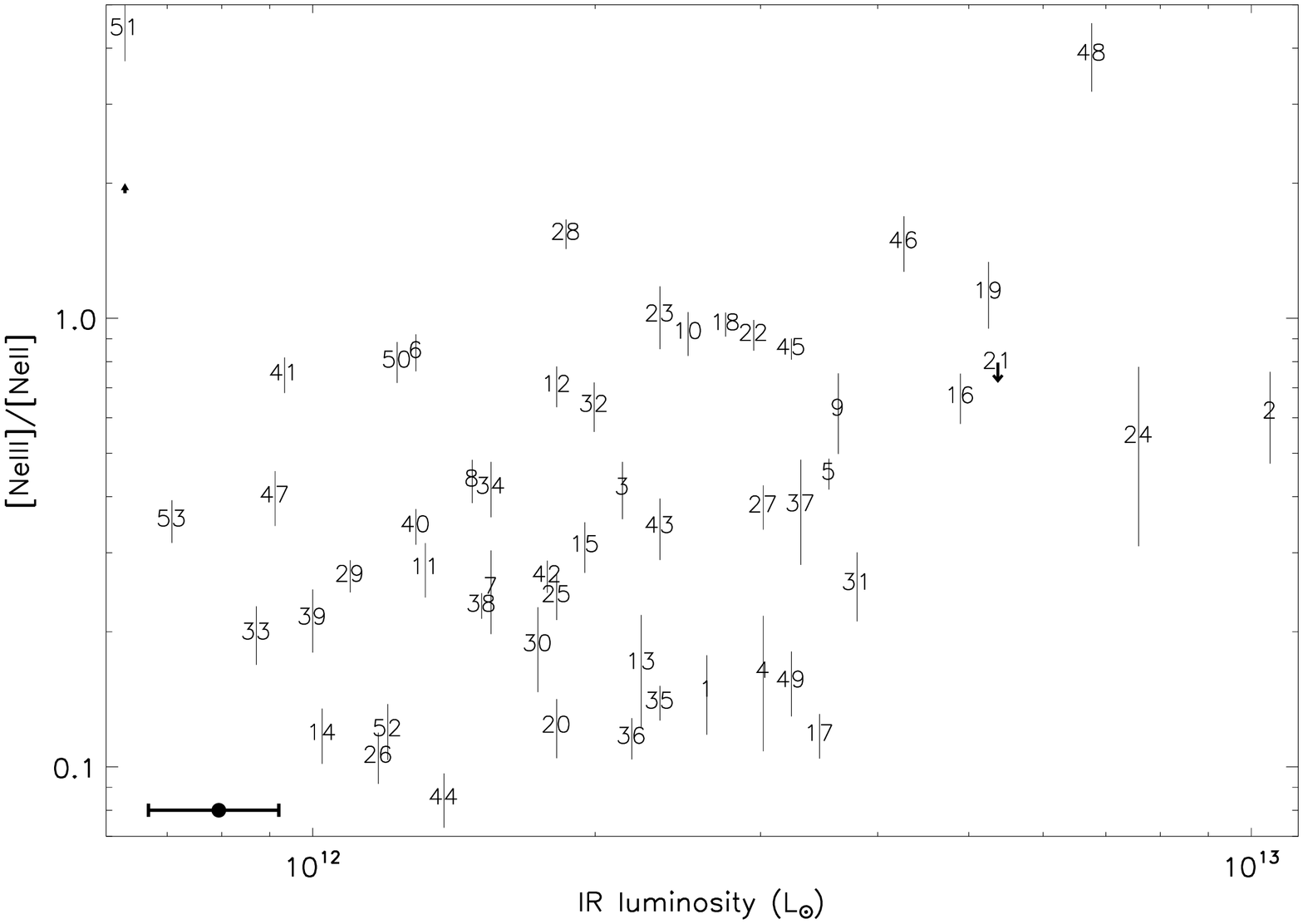}
\end{minipage}
\caption{[NeIII]$\lambda15.55$/[NeII]$\lambda12.81$ vs infrared luminosity. The small arrow indicates the effect 
on a points position if the V band extinction is increased by $A_{V}=30$. 
The horizontal bar on the bottom left indicates a 20\%\ error on the IR luminosity.  \label{excitelum}}
\end{figure}

\begin{figure}
\begin{minipage}{160mm}
\includegraphics[angle=90,width=150mm]{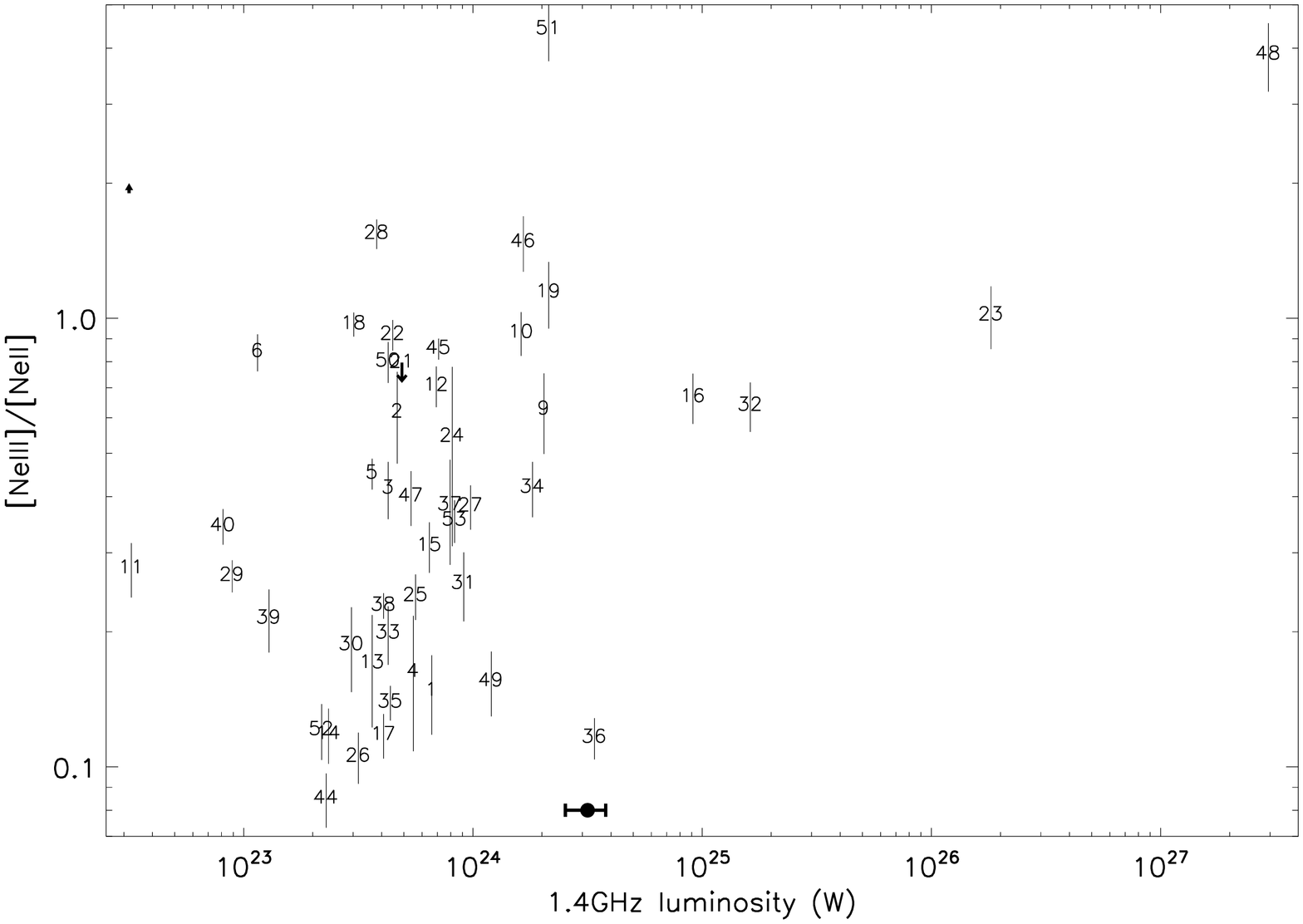}
\end{minipage}
\caption{[NeIII]$\lambda15.55$/[NeII]$\lambda12.81$ vs 1.4GHz luminosity. The small arrow indicates the effect 
on a points position if the V band extinction is increased by $A_{V}=30$. 
The horizontal bar on the bottom left indicates a 10\%\ error on the 1.4GHz luminosity. \label{excitelum2}}
\end{figure}

We can also explore variations in excitation as a function of the total emission. In Figure \ref{excitelum} we plot the [NeIII]/[NeII] ratio against infrared 
(rest-frame $1-1000\mu$m) luminosity. There is no discernible correlation. Two ULIRGs, Mrk 463E and 3C 273, have elevated [NeIII]/[NeII] ratios (though with very different IR luminosities), while the rest of the sample is confined to [NeIII]/[NeII] ratios in the range 0.08 - 1.05, irrespective of their IR luminosity. This picture remains broadly the same if we substitute IR luminosity for radio luminosity (Figure \ref{excitelum2}). In this case, objects with 1.4GHz luminosities lower than $\sim10^{24.5}$W show no discernible correlation between radio luminosity and [NeIII]/[NeII] ratio. Those objects with higher radio luminosities appear to be confined to higher [NeIII]/[NeII] ratios, and with a narrower spread, though the number of objects with L$_{1.4}>10^{24.5}$W is too small to draw firm conclusions. We conclude therefore that neither 1-1000$\mu$m luminosity or $1.4$GHz luminosity are good proxies for the excitation of the NLR in ULIRGs.

\subsection{Starburst \& AGN activity}
Mid-IR emission from galaxies can arise from five sources; (1) a non-thermal component, 
e.g. supernova remnants, (2) photospheres of evolved stars, (3) ionized gas, (4) dust grains, and (5) molecular gas. In most ULIRGs the latter three sources dominate the mid-IR emission. 
The ionized gas gives rise to the fine structure lines, while the smaller 
dust grains produce the mid-IR continuum longward of about 10$\mu$m. Larger dust grains give rise to features such as the absorption 
features at 9.7$\mu$m and 18$\mu$m \citep{leb79,roc85,chi06}. Large molecules give rise to several emission and absorption features; the most 
prominent emission features are seen from 3.3$\mu$m to about 19$\mu$m and arise from bending and stretching modes of Polycyclic Aromatic Hydrocarbons \citep{pug89,hon01}. 
Other molecular features include several Hydrogen emission lines, as well as weaker features such as CO, HCN and C$_{2}$H$_{2}$ gas absorption \citep{spo06,ber06,lah07}. 

The use of these mid-IR emission and absorption features as diagnostics of the power source in ULIRGs is however not straightforward. The fundamental 
problem is that dust and gas simply require a source of ionizing radiation to emit in the mid-IR, and do not particularly care what the source 
of that ionizing radiation is. In principle, a population of hot stars or an accretion disk around a supermassive black hole both serve equally well. 
Exhaustive reviews of these difficulties are given elsewhere (e.g. \citealt{lfs06,dal06}), so here we highlight some examples. Fine structure line fluxes 
are affected by several factors, particularly in starbursts, where the age of the burst, the geometry of the region and the upper and lower bounds on the IMF can all 
have a drastic effect on observed line ratios (e.g. fig 6 of \citealt{tho06}). The factors determining PAH strengths are still poorly understood; one factor that is known to have a 
significant effect is metallicity, with strong suppression of PAHs seen in low metallicity environments \citep{roc91,thu99,dwe05,wu06}. There is also no {\it a priori} reason why 
PAH bending and stretching modes cannot be excited by an AGN, although observationally PAHs seem to be almost exclusively associated with either star forming regions 
or ambient interstellar light \citep{pee04}. Significant silicate absorption on the other hand merely requires a mass of warm dust obscured by a significant 
column of cooler dust, and therefore on its own says nothing about what is heating the warm dust.

With these subleties in mind, we examine the nature of the starburst and AGN activity in our sample. We focus here on a limited number of diagnostics, deferring detailed analysis 
to upcoming papers. 

\subsubsection{Neon lines and star formation}\label{neosfr}
Recently, \citet{ho07} have suggested that the total luminosity of the [NeII]$\lambda$12.81 and [NeIII]$\lambda$15.56 lines are a diagnostic of star formation rates 
in galaxies across a wide range in IR luminosity. This method potentially offers a number of advantages over other mid-IR star formation diagnostics; the Neon lines in question are 
strong and easily observed, and the dependence on an accurate extinction correction is small. 
With ULIRGs however, there is the obvious caveat that some or all of the [NeII]$\lambda$12.81 and [NeIII]$\lambda$15.56 may originate in regions excited by an AGN rather than a starburst, 
so in this section we explore the use of the [NeII]$\lambda$12.81 and [NeIII]$\lambda$15.56 lines as star formation rate diagnostics in our sample. 

In Figure \ref{hoketo} we plot the [NeIII]+[NeII] luminosities of our sample against total IR luminosity, and overplot the relation given in equation 2 of \citet{ho07}. 
There is a large scatter in the [NeIII]+[NeII] luminosities as a function of IR luminosity for our sample, comparable to the scatter seen in figure 1 of \citet{ho07}, but we see 
a clear upward trend in [NeIII]+[NeII] luminosity with increasing IR luminosity. A power law fit yields $L_{N} \propto L_{ir}^{0.75}$, and a horizontal line fit (i.e. 
no dependence of [NeIII]+[NeII] luminosity on IR luminosity) is ruled out at $>5\sigma$ significance. The slope of the relation for the ULIRGs is consistent with the slope of the 
Ho \& Keto relation plotted in Figure \ref{hoketo}, but the ULIRGs are systematically lower, offset by $\sim0.4$ dex, on average. There are several possible explanations for this 
offset. First is that an AGN is contributing to the total IR luminosity but not to the Neon line fluxes, however this explanation seems unlikely, as we see systems 
with known IR-luminous AGN scattered on either side of the Ho \& Keto line; Mrk 463E and NGC 6240 both lie above it, whereas Mrk 231 and IRAS 03158+4227 (which contains a 
Compton-thick AGN, \citealt{ris00}) both lie well below it. Second is a different electron density in the NLR of ULIRGs; either significantly lower than that seen in lower 
luminosity starbursts, or much higher and approaching the critical 
density. A high electron density however is ruled out (see \S\ref{elecdens}), and there is no evidence for electron densities substantially lower 
than in lower luminosity systems. A significantly lower metallicity would also serve to lower Neon luminosities for a given IR luminosity, but again there is no 
evidence (or indeed plausible motivation) for such low metallicities in local ULIRGs. Finally, this offset could be caused by higher extinction. This explanation seems the 
most likely. Though the Ho \& Keto relation was derived without correcting for extinction, we expect ULIRGs to suffer heavier extinction in their nuclear regions 
relative to lower luminosity starbursts. The offset seen for ULIRGs corresponds to $A_{V}=40\pm20$ over and above 
the extinction seen in lower luminosity systems.

\subsubsection{PAHs and star formation}\label{pahsfr}
The correlation between [NeIII]$\lambda$15.56 + [NeII]$\lambda$12.81 luminosity and IR luminosity is consistent with the idea that there is a correlation between [NeIII]$\lambda$15.56 + [NeII]$\lambda$12.81 luminosity and star formation rate in ULIRGs. This on its own however does not support such a hypothesis, as we expect the luminosity of any mid-IR fine-structure line to correlate approximately with the mid-IR luminosity of the ionizing souce, irrespective of what that source may be. Therefore, to test this hypothesis further, we examine the relationship between [NeIII]$\lambda$15.56 + [NeII]$\lambda$12.81 luminosity and the luminosity of PAH features. 

The origin of PAH features and their use as star formation indicators are still controversial (e.g. \citealt{pee04,for04}). Observationally however, PAH features are usually prominent in starburst galaxies, with previous authors noting a good correlation between the strength of PAH features and IR luminosity in starbursts \citep{bra06}, but appear to be weak or absent in AGN \citep{wee05}. The cause of this dichotomy is thought to be a combination of two factors. First, the UV radiation field from an accretion disk around a black hole is harder than the UV radiation field from a starburst, and is therefore more adept at destroying the C-H and C-C bonds in PAHs. Second, a luminous AGN produces a prominent mid-IR continuum which can drown out PAH features, even if there is a vigorous starburst present \citep{lau00}.

Even though there is good evidence that PAHs originate in star forming regions, the behaviour of individual PAH features as a function of star formation rate remains uncertain, with strong variations observed in individual PAH feature strengths between different starburst galaxies \citep{jds07}. Therefore, we elect to use the combined luminosities of two PAH features in order to reduce the likely scatter introduced by variances in individual PAH strengths. The choice of the two PAH features is a straightforward one; the most luminous PAH features are those at 6.2$\mu$m, 7.7$\mu$m and 11.2$\mu$m, but the 7.7$\mu$m feature is difficult to measure as it lies in a crowded part of the spectrum. Therefore, even though the 11.2$\mu$m feature can be significantly affected by silicate absorption at 9.7$\mu$m, we use the luminosities of the PAH 6.2$\mu$m and 11.2$\mu$m features. These PAH luminosities are measured from the low resolution IRS spectra presented in \citet{des07}. For the 11.2$\mu$m PAH feature the luminosity was measured by integrating over 10.8$\mu$m - 11.8$\mu$m in the continuum-subtracted spectra, with the equivalent width evaluated at 11.25$\mu$m. The relevant parameters for the 6.2$\mu$m PAH feature were 5.90$\mu$m - 6.55$\mu$m and 6.22$\mu$m, respectively. 

In Figure \ref{hoketoalt} we plot [NeIII]+[NeII] luminosity against the combined luminosities of the PAH 6.2$\mu$m and 11.2$\mu$m features\footnote{A similar plot but using only the 6.2$\mu$m luminosity or the 11.2$\mu$m luminosity yield plots with significantly greater scatter}. The Neon line luminosities and PAH luminosities clearly track each other, lending support to the idea that, on average, lower ionization Neon lines originate in regions heated mostly by star formation. A power law fit yields:

\begin{equation}
L_{N} = (0.17^{+0.46}_{-0.12})\times L_{P}^{1.02\pm0.05} \label{neopah}
\end{equation}

\noindent where L$_{N}$ and L$_{P}$ are the [NeIII]+[NeII] and 6.2$\mu$m+11.2$\mu$m luminosities (in any units, as long as they are both the same), respectively. Equation \ref{neopah} 
is solely intended to indicate the trend, and evidently is not a good physical model for the data. It is notable however that the scaling between [NeIII]+[NeII] luminosity and PAH 6.2$\mu$m+11.2$\mu$m luminosity predicted from this fit is, to within the error on the exponent of $L_{P}$, linear. Interestingly, we see a comparable slope (though different normalization) if we instead plot [SIII]$\lambda18.713$ luminosity against PAH 6.2$\mu$m+11.2$\mu$m luminosity. This is consistent with the idea that the sizes of the lower ionization Neon and Sulfur emiting regions, and PAH emitting regions in ULIRGs scale linearly with each other with increasing luminosity, though a luminosity dependence on metallicity and/or electron density could also play a role.  

It is also straightforward to derive a relation between PAH 6.2$\mu$m+11.2$\mu$m luminosity and star formation rate, using equation 12 from \citet{ho07} combined with Equation 
\ref{neopah}. In doing so however, we use the updated conversion between Lyman continuum flux and star formation rate given by \citet{hir03}, rather than the relation in \citet{ken98}. 
By combining equation 12 from \citet{ho07} with the relations between Lyman continuum flux and star formation rate given in \S2.1 of \citet{hir03}, we obtain:

\begin{equation}\label{bigeqn}
SFR [M_{\odot} yr^{-1}] = 2.69\times10^{-41}\frac{L_{N} [ergs\ s^{-1}]}{f_{ion}(f_{Ne^{+}} + 1.67f_{Ne^{++}})}
\end{equation}

\noindent where SFR is the star formation rate, $L_{Neon}$ is the combined luminosity of the two Neon lines, $f_{ion}$ 
is the fraction of photons that actually contribute to ionizing the gas, and $f_{Ne^{+}}$ and $f_{Ne^{++}}$ are the fractional 
abundances of [NeII] and [NeIII], respectively. For ULIRGs however there is an additional complication. The conversion in 
\citet{hir03} assumes continuous star formation over $\gtrsim10^{8}$ years. Star formation in ULIRGs however is likely to 
occur in some form of `burst' over timescales of $10^{7}-10^{8}$ years. This means that the star formation rates in ULIRGs 
will be underestimated by a significant amount if the conversion in Equation \ref{bigeqn} is used unmodified. Following 
previous work \citep{ken98,tho06}, we adopt an {\it approximate} upward scaling of 50\% to correct for this difference. 
This scaling is likely to be sufficient for considering trends, but should be adjusted for specific objects where the age 
and nature of the burst are known. Assuming this scaling, plus linear scaling between Neon and PAH luminosity, 
$f_{ion}=0.6$, $f_{Ne^{+}}=0.75$ and $f_{Ne^{++}}=0.1$ \citep{ho07}, we arrive at:

\begin{equation}
SFR [M_{\odot} yr^{-1}] = 1.18\times10^{-41}L_{P} [ergs\ s^{-1}]\label{pahstar}
\end{equation}

\noindent where $L_{P}$ is in units of ergs s$^{-1}$. This assumes Solar metallicity and a Salpeter IMF spanning 
0.1M$_{\odot}$ to 100M$_{\odot}$, and is only applicable to objects where the PAH and Neon luminosities arise from 
recent star formation. 

The errors on the star formation rates derived using Equation \ref{pahstar} 
are difficult to quantify. The conversion from Ly$\alpha$ continuum flux to star formation rate has 
an error of $\sim$30$\%$ \citep{ken98}, and there is a significant extra error introduced by the scatter 
in Figure \ref{hoketoalt}. We therefore estimate that the errors on star formation rates derived using Equation 
\ref{pahstar} are of order 30\% for population studies, rising to at least 50\% for individual objects. Equation 
\ref{pahstar} does however offer the advantage that it does not require an accurate measure of the rest-frame 
1-1000$\mu$m luminosity, and hence is particularly suited to population studies at high redshifts. Equation 
\ref{pahstar} also gives plausible star formation rates for our sample. We defer a complete analysis 
to an upcoming paper, but note that for Arp 220 and using the PAH luminosities in \citet{des07}, Equation 
\ref{pahstar} gives a star formation rate of 57M$_{\odot}$ yr$^{-1}$, which is consistent with the total star 
formation rate inferred from radio observations, which give 50-100M$_{\odot}$ yr$^{-1}$ (\citealt{smi1}, but 
see also \citealt{par07}).

\subsubsection{Starburst vs. AGN diagnostics}
To distinguish between starburst and AGN power behind the 1-1000$\mu$m emission in ULIRGs, we need diagnostics that are sensitive to the observable differences between young stars, 
and an accretion disk around a supermassive black hole. For our purposes, there are two such differences. First, the ionizing radiation from an AGN is harder and (potentially) more 
intense than that from a starburst. Second, AGN occupy a smaller volume, $\sim0.1$pc, as opposed to a few tens to a few hundreds of pc for a starburst. In principle therefore, we might 
expect the averaged mid-IR spectra of AGN to exhibit three differences compared to those of starbursts; higher ionization fine structure lines, increased 
quantities of `hot' ($\gtrsim100$K) dust, and (potentially) increased obscuration. We explore each of these differences in this section. 

We start with the simplest possible diagnostic; the detection (or otherwise) of individual lines. For examining starburst and AGN activity, the [NeV]$\lambda$14.32 
and [OIV]$\lambda$25.89 lines are the most useful single line diagnostics. Both lines can be strong in planetary nebulae and young supernova remnants \citep{oli99}, and [OIV] is
sometimes seen in W-R star spectra. Both lines are however weak in spectra of star forming regions (e.g. \citet{lut98}), while being strong in spectra of AGN.
As [NeV]$\lambda$14.32 has $E_{ion}=97.1$ eV, compared to 54.9eV for [OIV]$\lambda$25.89, and because models indicate that [NeV]$\lambda$14.32 is unlikely to be strong in 
galaxies without an AGN \citep{voi92}, we focus most of our attention on the [NeV]$\lambda$14.32 line. 

The presence of [NeV]$\lambda$14.32 cannot provide meaningful constraints on the bolometric luminosity 
of an AGN, simply becase a comparable [NeV]$\lambda$14.32 flux can be produced by a faint AGN with low 
obscuration, or a luminous AGN with high obscuration. From Table \ref{linefluxa} however we can see 
that [NeV]$\lambda$14.32 is detected in 22 out of 53 objects, providing direct spectroscopic evidence for 
an AGN in these 22 objects. Our sample is flux- rather than volume-limited, hence we cannot draw firm conclusions about 
local ULIRGs as a whole, however this suggests that an AGN provides a non-negligible fraction of the mid-IR flux in $\sim42$\% of local ULIRGs. 
This is significantly higher than the fraction of ULIRGs that show spectroscopic signatures of 
AGN activity in their optical or near-IR spectra, $20\% - 25\%$ \citep{vei2}, though strictly speaking 
the \citealt{vei2} sample and our sample are not directly comparable as their selections are different. 
Interestingly, we see no 
convincing trend  in the detection of [NeV]$\lambda$14.32 as a function of IR luminosity. From Table 
\ref{linefluxa} the ULIRGs with a [NeV]$\lambda$14.32 detection are fairly evenly spread between IR 
luminosities of $10^{11.80}$L$_{\odot}$ and $10^{12.88}$L$_{\odot}$; 10/22 ULIRGs with a [NeV]$\lambda$14.32 
detection have L$_{ir}>10^{12.38}$L$_{\odot}$, while 20/53 of the whole sample have 
L$_{ir}>10^{12.38}$L$_{\odot}$. This contrasts with optical/near-IR spectroscopic surveys, which 
report a rising fraction of ULIRGs with AGN signatures, reaching $35\%-50\%$ at 
$>10^{12.38}$L$_{\odot}$ \citep{vei2}.\footnote{the `boundary' luminosity quoted by \citealt{vei2} 
is $10^{12.30}$L$_{\odot}$, however this is for $H_{0}=75$ km s$^{-1}$ Mpc$^{-1}$ and $q_{0} = 0.0$} 
A plausible reason for this discrepancy is high levels of extinction toward the nuclei of ULIRGs. From 
\S\ref{neosfr}, the increase in extinction toward the nuclei of ULIRGs compared to lower luminosity 
systems is of the order $A_{V}\simeq40$, making it harder to see AGN signatures in the optical 
or near-IR. Mid-infrared spectroscopy on the other hand appears to be a sensitive probe of the presence 
or otherwise of AGN in IR-luminous systems. We detect [OIV]$\lambda$25.89 in 21/22 objects that show 
[NeV]$\lambda$14.32, and detect [OIV]$\lambda$25.89 in only two objects that do not show [NeV]$\lambda$14.32, 
suggesting that [OIV]$\lambda$25.89 is a good, but not perfect proxy for [NeV]$\lambda$14.32. 

We next consider diagnostics based on fine-structure line ratios. In Figures \ref{hiresdiaga} and \ref{hiresdiagb} 
we plot [NeV]/[NeII] and [OIV]/[NeII] against IR luminosity, together with the predicted AGN contribution to the total IR luminosity based on these line ratios \citep{stu02}. 
Both diagrams predict a broad spread in power source, ranging from 100\% starburst to 100\% AGN. In both diagrams however only 10-12 ULIRGs are above the 40\% AGN 
line, and more than half the sample lie below the 20\% lines. This is consistent with star formation being the dominant contributor to the IR emission in most ULIRGs, 
with only 20\% of ULIRGs hosting an AGN with a comparable or greater IR luminosity than the starburst, in agreement with studies at other wavelengths 
\citep{vei2,far4,fra03}. There are however many caveats in using such simple diagnostics \citep{stu02}, and Figures \ref{hiresdiaga} and \ref{hiresdiagb} should 
not be considered reliable diagnostics of the power source in individual ULIRGs, but rather as crude indicators of trends. It is however notable that we seem to 
have results consistent with other diagnostics for (some) individual objects. For example, both diagrams predict a small contribution from an AGN in Arp 220, consistent with 
results from the X-ray \citep{cle02} and from ISO \citep{stu96}, and 3C 273, Mrk 1014, Mrk 463E and IRAS 05189-2524 are all predicted to contain a luminous AGN, in line with 
previous work \citep{sha76,maz91,bol02,far05}.

Given the likely origin of the [NeV]$\lambda$14.32 line in AGN, we can use the [NeV]/[NeII] ratio to test other mid-IR AGN diagnostics. 
One such diagnostic is the IRAS 25$\mu$m/60$\mu$m flux density ratio; it has been suggested that `warm' objects (those with 
$f_{25}/f_{60}>0.2$) are more likely to contain an AGN than `cool' ($f_{25}/f_{60}<0.2$) objects \citep{deg85,san2}. In Figure 
\ref{nevcfwarm} we plot [NeV]/[NeII] against $f_{25}/f_{60}$. Just over half of the `warm' objects have [NeV]$\lambda$14.32 detections, compared 
to about one third of the `cool' objects. Furthermore, for those objects with [NeV]$\lambda$14.32 detections, the `warm' objects have 
systematically higher [NeV]/[NeII] ratios than the cool objects. While our small sample sizes (particularly for the warm objects) render 
any conclusions tentative, we infer from this that the IRAS 25$\mu$m/60$\mu$m flux density ratio is a reasonably good, though not perfect 
diagnostic for the presence of an IR-luminous AGN in a ULIRG. 

We move on to consider diagrams with a diagnostic of the starburst and AGN luminosities on both the $x$ and $y$ axes. A natural diagnostic to combine with a fine-structure line 
ratio is a measure of the strength of the PAH features. To measure PAH strength we use the equivalent width (EW) of the 6.2$\mu$m PAH feature, taken from \citet{des07}, as this feature is strong, and lies in a 
relatively uncluttered part of the mid-IR spectra of ULIRGs, though its use does depend on an accurate correction for water ice and/or aliphatic hydrocarbon absorption. A mid-IR fine-structure line ratio $vs.$ PAH EW diagnostic diagram has been used by several authors for many classes of IR-luminous galaxy (e.g. 
\citealt{gen,stu02}). 

In Figures \ref{hipahdiaga} and \ref{hipahdiagb} we plot [NeV]/[NeII] and [OIV]/[NeII] against the equivalent width of the 6.2$\mu$m PAH 
feature, together with the linear mixing ratios from \citet{stu02} and \citet{arm06}. Here we again see a broad spread in fractional AGN 
and starburst luminosities, from `pure' AGN to `pure' starbursts. Those sources classified by Figures \ref{hipahdiaga} and \ref{hipahdiagb} 
as AGN dominated tend to have PAH$^{6.2\mu}_{EW} \lesssim 0.05$, [NeV]/[NeII]$\gtrsim 0.2$ and [OIV]/[NeII]$\gtrsim 0.5$, whereas the `pure' 
starbursts tend to have PAH$^{6.2\mu}_{EW} \gtrsim 0.2$, [NeV]/[NeII]$\lesssim 0.09$ and [OIV]/[NeII]$\lesssim 0.2$. The IR emission from 
the majority of the sample are still predicted to be powered mainly by star formation, though this trend is not so clear as from Figure 
\ref{hiresdiaga}. We see broad (to within a factor of three) agreement between the two axes, and also that no object lies in the `forbidden' 
top right hand corner of either diagram. In both diagrams, the six objects which show a detection of the [ClII]$\lambda$14.37 line 
($\S$\ref{rarelines}) are mostly located toward the right hand side, consistent with the postulated origin of this line in star forming regions. 

It is however notable that several sources lie in a `forbidden' region in each diagram, where the PAH equivalent widths suggest a low starburst 
contribution, but the fine-structure line ratios suggest a low AGN contribution. There are, broadly, four possible reasons for this 
discrepancy\footnote{A fifth possibility is variation in the filling factor of the coronal-line region clouds. A low filling factor would 
imply a small ionization parameter $U$, and would therefore give an unusually low [NeV]/[NeII] ratio, for example. We do not however have 
the data to address this possibility}; (1) underestimated 6.2$\mu$m PAH EWs due to nearby water ice and/or aliphatic hydrocarbon absorption,
(2) destruction of PAHs in {\it luminous} starburst environments \citep{rig,far4,bei06}, (3) suppression of PAH EWs in {\it compact} starburst 
environments, and (4) increased general obscuration levels. We do not consider the first two possibilities likely though. We have in all cases 
used ice absorption-corrected values for the 6.2$\mu$m PAH EWs (the correction is of order $10\%$ for the 6.2$\mu$m feature and less than $5\%$ 
for the 11.2$\mu$m feature, \citealt{spo07}) and it is unlikely that any residual ice absorption could shift sources by such a significant distance 
on this plot. If the PAHs were being destroyed in extremely luminous starbursts then we would expect the objects in the `forbidden' region to all 
have high IR luminosities, but the objects in this region (e.g. IRAS 08572+3915, IRAS 00397-1312, IRAS 15462-0450) span a wide range in IR 
luminosity. A combination of increased overall obscuration and increased starburst compactness therefore seem the most likely culprits. 
The mixing ratios in both diagrams are based on starbursts and AGN with IR luminosities of $10^{10} - 10^{11}$L$_{\odot}$. Their use therefore 
assumes a linear scaling between starburst and AGN luminosity, and overall obscuration and starburst geometry, to give rise to comparable observed 
[NeV]/[NeII] and [OIV]/[NeII] ratios and PAH 6.2$\mu$m EWs. These assumptions are unlikely to be valid. For example, previous authors have noted 
$A_{\lambda}$'s of ten or more, even in the mid-IR \citep{gen}, and we derived $A_{V}\simeq40$ over and above lower luminosity stabursts in 
\S\ref{neosfr}. Increased overall obscuration will lead to greater suppression of the [NeV] and [OIV] lines compared to the [NeII] line, as 
these regions must lie closer to the starburst and/or AGN. This increased obscuration will also arise in lower than expected PAH luminosities, 
but is unlikely to affect their EWs. However, we expect more compact starbursts to have a stronger mid-IR continuum, which will lead to 
suppression of PAH EWs. Overall therefore, increased overall obscuration combined with more spatially compact starbursts in ULIRGs 
compared to lower luminosity systems can plausibly explain the outliers in Figures \ref{hipahdiaga} and \ref{hipahdiagb}. 

An important caveat however is that diagrams like Figures \ref{hipahdiaga} and \ref{hipahdiagb} are not suited to studying the physics of 
individual objects, and can often get the answers wrong. For example, if the power source is so obscured that it does not emit 
significantly at wavelengths shortward of $\sim25\mu$m, then mid-IR fine structure line ratio $vs.$ PAH EW diagnostic plots can give misleading 
results \citep{pee04}. Some `outliers' are therefore understandable. One example is IRAS 15206+3342 (\#28). This object lies squarely in the `starburst' part 
of both plots, and from this one might conclude that it does not contain an energetically significant AGN. Its UV spectrum \citep{far05} 
however shows clear evidence for a broad absorption line QSO. A second example is NGC 6240, which also lies in the `starburst' part of 
both plots, despite the extensive evidence for an IR-luminous AGN in this source \citep{arm06a}. The key point is that diagnostics such 
as those in Figures \ref{hipahdiaga} and \ref{hipahdiagb} probe the physics of specific environments within ULIRGs. To obtain a complete 
picture of an individual object therefore requires multiple diagnostics to probe multiple environments.

\subsubsection{Silicate absorption} \label{silpower}
We explore the issue of mid-IR obscuration in ULIRGs further by combining the diagnostics discussed previously with a measure of the strength of the 9.7$\mu$m silicate 
absorption feature, $S_{sil}$:

\begin{equation}
S_{sil} = ln\left(  \frac{F_{obs}(9.7\mu m)}{F_{cont}(9.7\mu m)} \right)\label{silstrength}
\end{equation}

\noindent where $F_{obs}$ is the observed flux at rest-frame 9.7$\mu$m, and $F_{cont}$ is the underlying continuum flux at rest-frame 9.7$\mu$m deduced 
from a spline fit to the continuum flux at rest frame 5.0-7.0, 14.0-14.5, and 25.0-40.0$\mu$m. A complete description of the method used to measure 
$S_{sil}$ can be found in \citealt{spo06} and \citealt{lev07}. The silicate strengths for our sample are measured from the low resolution spectra, and 
are presented in \citet{spo07}.  

In Figure \ref{oivsi} we plot $S_{sil}$ against the [OIV]/[NeII] line ratio, along with the `reference' starbursts and AGN 
\citep{bra06,wee05}. The reference starbursts and AGN separate well on the $y$ axis, 
with most of the AGN above [OIV]/[NeII]=0.2 and most of the starbursts below. Nearly all the starbursts and AGN are confined to $-0.5 < S_{sil} < 2.0$. The ULIRGs are 
found over the whole range of [OIV]/[NeII] ratios where the reference samples are seen, but are offset on the $x$ axis, with greater 
silicate strengths for a given [OIV]/[NeII] ratio. The simplest interpretation of this is that the starbursts in ULIRGs are similar in nature to those in lower luminosity 
systems, but with significantly greater ($A_{V}$'s of a few tens) total obscuration, consistent with the conclusion in \S\ref{neosfr}.

It would be natural, from this, to postulate that silicate strength is a measure of the total obscuration of the starburst. If so, then we 
might expect a correlation between star formation rate and silicate strength; the reasoning being that a more luminous starburst will have a deeper 9.7$\mu$m absorption feature as the starburst has an increased total dust column. It is straightforward to test this hypothesis; in Figure \ref{sidepthcode} we duplicate Figure \ref{hoketoalt}, but this time color-coded each object according to its silicate strength. Clearly, there is no correlation. Objects with the strongest silicate absorption seem to be confined to the lower left 
of the plot, though with only three objects it is impossible to draw firm conclusions. The other objects seem to be scattered randomly; strongly absorbed and weakly absorbed sources are found from the top right to the bottom left. Furthermore, if $S_{sil}$ was simply a measure of the obscuration of the starburst, then we would expect the sources with the greatest negative offset from the Ho \& Keto line in Figure \ref{hoketoalt} to have the largest values of $S_{sil}$, but this is not the case. For example, three of the sources with the greatest negative offset in Figure \ref{hoketoalt}, IRAS 06035-7102, IRAS 20551-4250, and 3C 273, range from having a deep silicate absorption feature to a silicate emission feature. We infer from this that silicate strength is not simply a measure of the total obscuration of the starbursts in ULIRGs. A similar conclusion was reached by \citet{hig06}, based on the molecular Hydrogen line strengths.  

We move on to consider a different scenario, in which the 9.7$\mu$m silicate absorption in ULIRGs is affected by the AGN as well as, or instead of, the starburst. To examine 
this possibility we duplicate the plot in Figure \ref{hipahdiagb} in Figure \ref{sidepthcodeb}, again color-coding each point by its silicate strength. From 
this we see that both strongly and weakly absorbed sources are found across the span of the plot, and there is no appreciable trend in {\it average} silicate 
strength in any direction. There is however one interesting trend. Those sources towards the left side of the diagram tend to have either very deep absorption, 
or very shallow to no absorption, whereas those objects on the right hand side of the plot seem to be almost universally `averagely' absorbed. Of the 13 objects 
on the right hand side, all but two (IRAS 15206+3342 \& IRAS 23128-5919) have $0.8<S_{sil}<2.4$. Of the 13 objects on the left hand side, only one 
(IRAS 12514+1027) does not satisfy $S_{sil}>2.4$ or $S_{sil}<0.8$. Overall, there is a clear shift in the distribution of values of $S_{sil}$ as we 
move from the left hand side to the right hand side of the plot, from a single, broad peak centered at around $S_{sil}\simeq1.5$ to two narrower peaks, 
one at $S_{sil}\simeq2.8$ and one at $S_{sil}\simeq0.2$. This shift is conceptually similar to the `fork' diagram in \citet{spo06}.

These distributions in silicate strength can be interpreted in one of two ways. The first is that very strong or very weak silicate absorption indicates the presence of an AGN, with 
moderate silicate absorption indicating the presence of a starburst. For an AGN we expect a strong dependence of observed properties on viewing angle, as nearly all lines of evidence suggest that the dust in AGN is arranged in a planar structure, whether that structure is a torus \citep{sch05}, a flared disk \citep{fri06}, or discrete clouds \citep{eli06}. For a starburst however we expect a weaker (though still possibly significant) dependence on viewing angle; millimeter interferometry has shown that the starburst regions in ULIRGs are dense and compact \citep{tac06}, while recent theoretical (e.g. \citet{sil05}) and indirect observational evidence suggests that starbursts in ULIRGs are unlikely to have a disklike structure \citep{fis06}.  We therefore expect that the distribution of values of $S_{sil}$ in obscured AGN will be bimodal, with a high apparent obscuration when the AGN is viewed edge on, but low obscuration when the AGN is viewed face on, but that the dependence of $S_{sil}$ on the viewing angle of the starburst will be weaker. This is supported by the fact that optical QSOs usually show silicates in emission \citep{hao06}, and that one of the two objects on the right hand side of the lower panel of Figure \ref{sidepthcode} is IRAS 15206+3342 which, as previously mentioned, contains a BAL QSO. 

The second way is that the sources with $S_{sil}\gtrsim2.4$ instead contain an extremely obscured starburst, and that only those sources with either a [NeV] detection or a silicate feature in emission contain an IR-luminous AGN. This scenario is plausible as a very compact, highly obscured starburst could also result in suppression of PAH features, and would also explain the absence of [NeV] detections (if the starburst surrounds the AGN) in the few objects on the left hand side of Figure \ref{sidepthcodeb} with $S_{sil}\gtrsim2.4$. With the data available to us it is difficult to choose between these two possibilities. Both scenarios are consistent with the  `fork' diagram in \citet{spo06} for example. With some reserve therefore, we propose that the first scenario is more likely, and that moderate silicate absorption signifies the presence of a dominant starburst, but that very deep or very shallow silicate absorption in ULIRGs likely signifies the presence of a bolometrically significant AGN.

\section{Conclusions}
We have presented an atlas of fine structure lines and other emission features measured from high resolution mid-infrared spectra of 53 ULIRGs at 0.018$<z<$0.319, 
taken using the Infrared Spectrograph onboard the Spitzer space telescope. We have employed a variety of diagnostics using these emission lines as well as those 
based on PAH features and the strength of the 9.7$\mu$m silicate absorption feature to investigate the power source behind the infrared emission. Our conclusions are:

\noindent 1) All of the spectra show various of fine structure emission lines of Neon, Oxygen, and Sulfur, as well as one or more molecular Hydrogen lines. 
We see the [NeII]$\lambda$12.81, [NeIII]$\lambda$15.56 and [SIII]$\lambda$18.71 in most of the sample, and [SIV]$\lambda$10.51 in just under half the sample. The 
higher ionization lines [NeV]$\lambda$14.32 and [OIV]$\lambda$25.89 are detected in just under half the sample, while [NeV]$\lambda$24.32 is detected in about one 
third of the sample. Rarer lines include [ClII]$\lambda$14.37 (six objects), HI 7-6$\lambda$12.37 (four objects), [PIII]$\lambda$17.89 (two objects), and [ArV]$\lambda$13.10 
(two objects). Some objects also show low-ionization iron lines, including [FeII]$\lambda\lambda\lambda$17.94,24.52,25.99 and [FeIII]$\lambda$22.93. The detection 
of three further lines is dependent on redshift, but we see [SIII]$\lambda$33.48 in $\sim$80\% of the objects where this line lies within the  
bandpass, [ArII]$\lambda$8.99 in just over $50\%$, and [SiII]$\lambda$34.82 in just over $50\%$.

\noindent 2) The presence of the [NeV]$\lambda$14.32$\mu$m line in 22/53 objects is direct spectroscopic evidence for the presence of an AGN that provides a significant, 
though not necessarily dominant fraction of the mid-IR flux in $\sim42\%$ of ULIRGs. Based on this, we find that the IRAS 25$\mu$m/60$\mu$m flux density ratio is a 
reasonable, though not perfect, diagnostic for the presence of an IR-luminous AGN in ULIRGs. In most, but not all objects where we see [NeV]$\lambda$14.32$\mu$m we also see 
[OIV]$\lambda$25.89, suggesting that [OIV]$\lambda$25.89 is a good proxy for the [NeV]$\lambda$14.32$\mu$m. In contrast, we see [SIV]$\lambda$10.51 in a 
surprisingly low fraction of the sample, given that its ionization energy is comparable to that of [NeIII]$\lambda$15.56, and that the H$_{2}$,S(3)$\lambda$9.66 line is seen in 
most of our sample. The most likely reason for this is that the the Neon and Sulfur emitting zones in our samples lie within regions that are more strongly extincted by 
silicate dust than the H$_{2}$ emitting regions; the increased extinction due to the nearby 9.7$\mu$m absorption feature would then weaken the apparent flux of 
[SIV]$\lambda$10.51 relative to [NeIII]$\lambda$15.56.

\noindent 3) We use the [NeIII]/[NeII] vs [SIV]/[SIII] plane to show that the excitation levels in the mid-IR emitting regions span more than two orders of magnitude 
in both the Neon and Sulfur line ratios. The range in both line ratios is comparable to that seen in starbursts and AGN with IR luminosities in the range 
$10^{10}<$L$_{ir}$(L$_{\odot})<10^{11.5}$, but we see a systematically lower [NeIII]/[NeII] ratio for a given [SIV]/[SIII] in our sample compared to systems with 
IR luminosities of $<10^{10}$L$_{\odot}$, possibly due to the increased density of gas in the NLR of ULIRGs. We use the 
[NeV]14.32$\mu$m/[NeV]24.32$\mu$m and the [SIII]18.71$\mu$m/[SIII]33.48$\mu$m line ratios to show that the electron densities in the mid-IR emitting regions of 
ULIRGs are $<10^{4}$ cm$^{-3}$ in all cases, well below the critical densities. 

\begin{figure}
\begin{minipage}{180mm}
\includegraphics[angle=90,width=160mm]{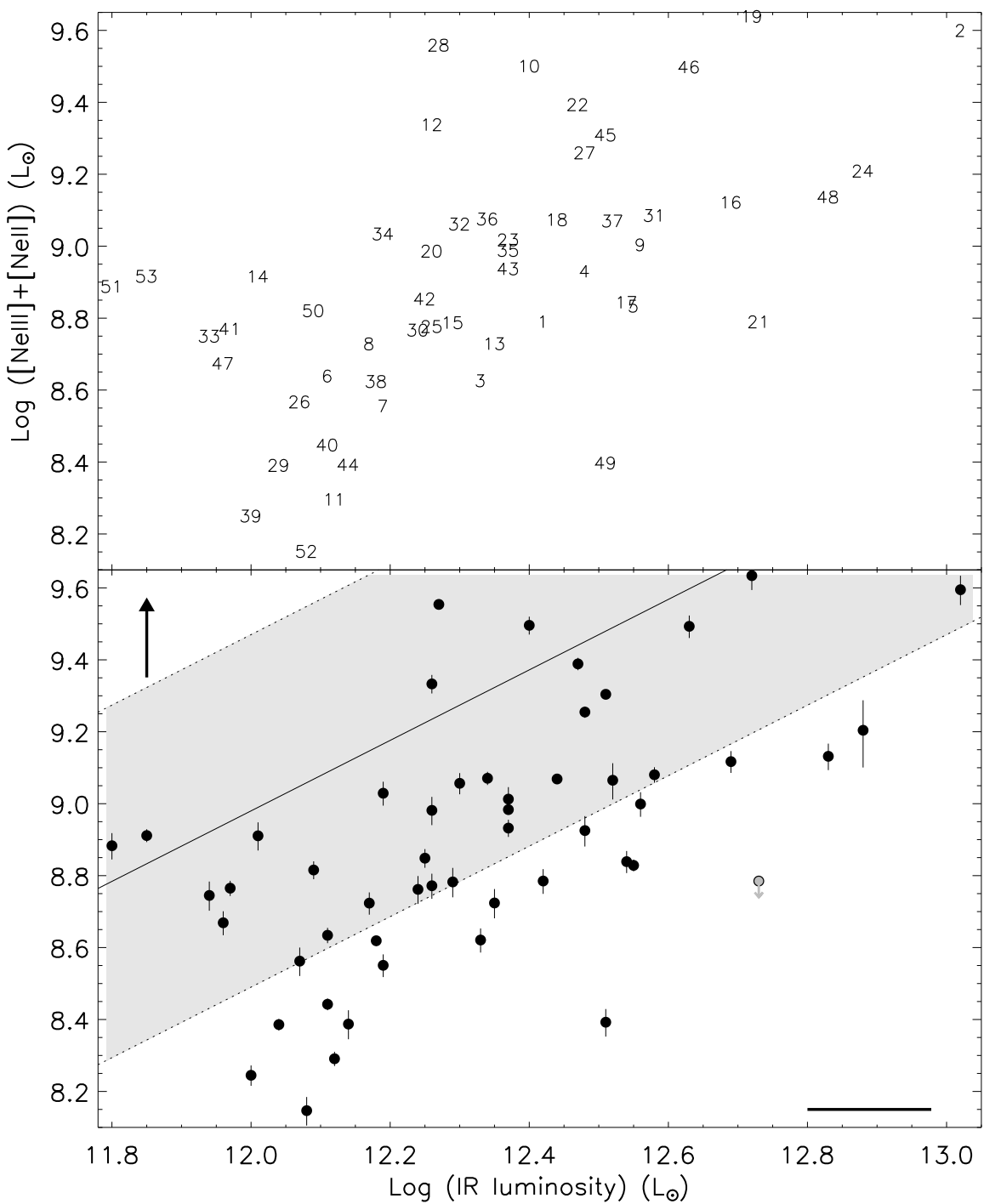}
\end{minipage}
\caption{Total luminosity of the [NeIII]$\lambda$15.55 and [NeII]$\lambda$12.81 lines vs IR luminosity. The solid 
line is the relation in \citet{ho07}, with the dashed lines indicating their 1$\sigma$ errors. ULIRGs with detections 
on both axes are plotted in black, while ULIRGs with limits on one or both axes are plotted in grey. The arrow indicates 
the effect on a points position if the V band extinction is decreased by $A_{V}=30$. The horizontal bar on the bottom 
left indicates a 20\%\ error on the IR luminosity.\label{hoketo}}
\end{figure}

\begin{figure}
\begin{minipage}{180mm}
\includegraphics[angle=90,width=160mm]{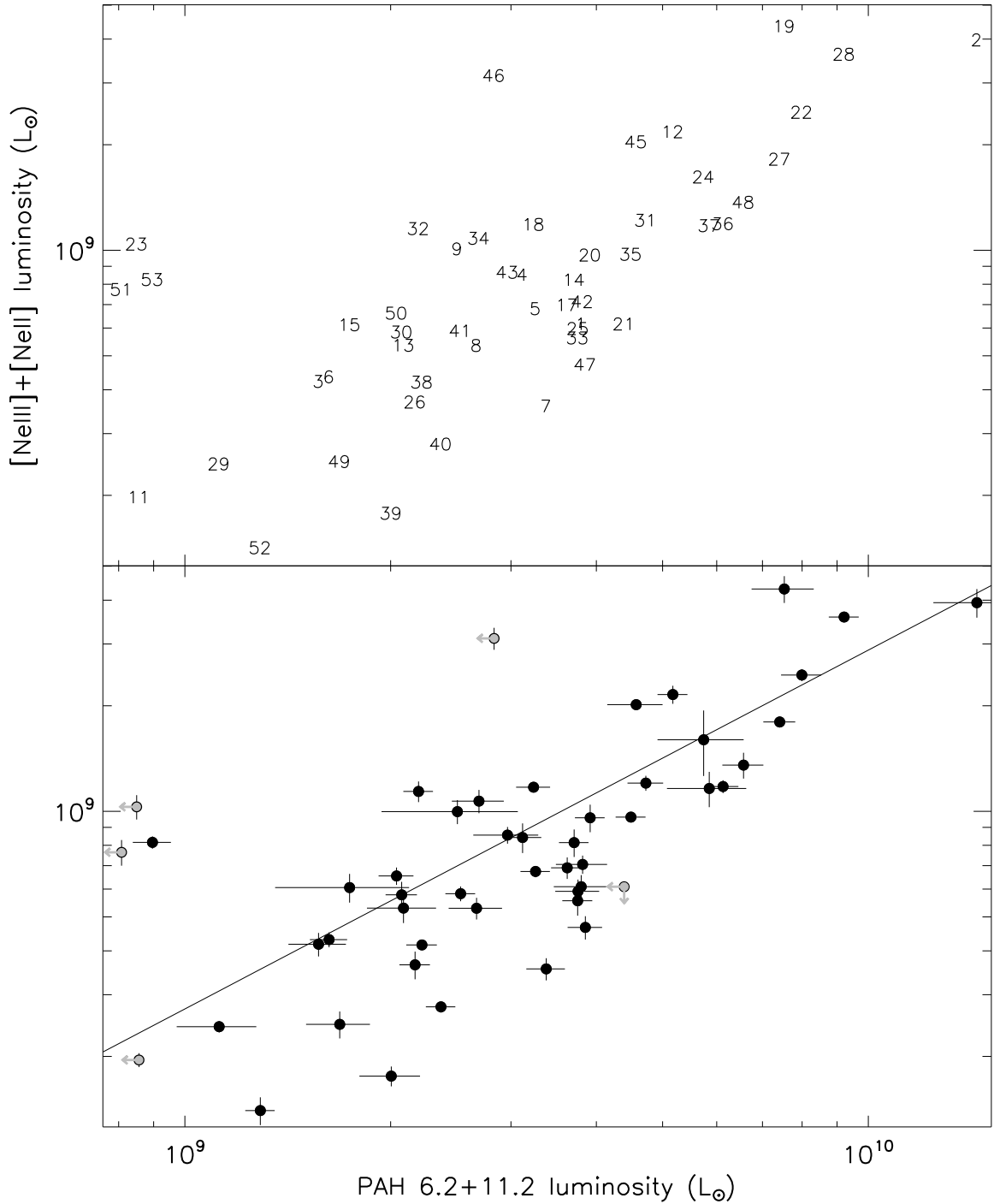}
\end{minipage}
\caption{The total luminosity of the [NeIII]$\lambda$15.55 and [NeII]$\lambda$12.81 lines vs the total luminosity of the PAH 6.2$\mu$m 
and 11.2$\mu$m features. ULIRGs with detections on both axes are plotted in black, while ULIRGs with limits on one or both axes are 
plotted in grey. The solid line indicates the best fit to the data, given in Equation \ref{neopah}. 
\label{hoketoalt}}  
\end{figure}

\begin{figure}
\begin{minipage}{170mm}
\includegraphics[angle=90,width=150mm]{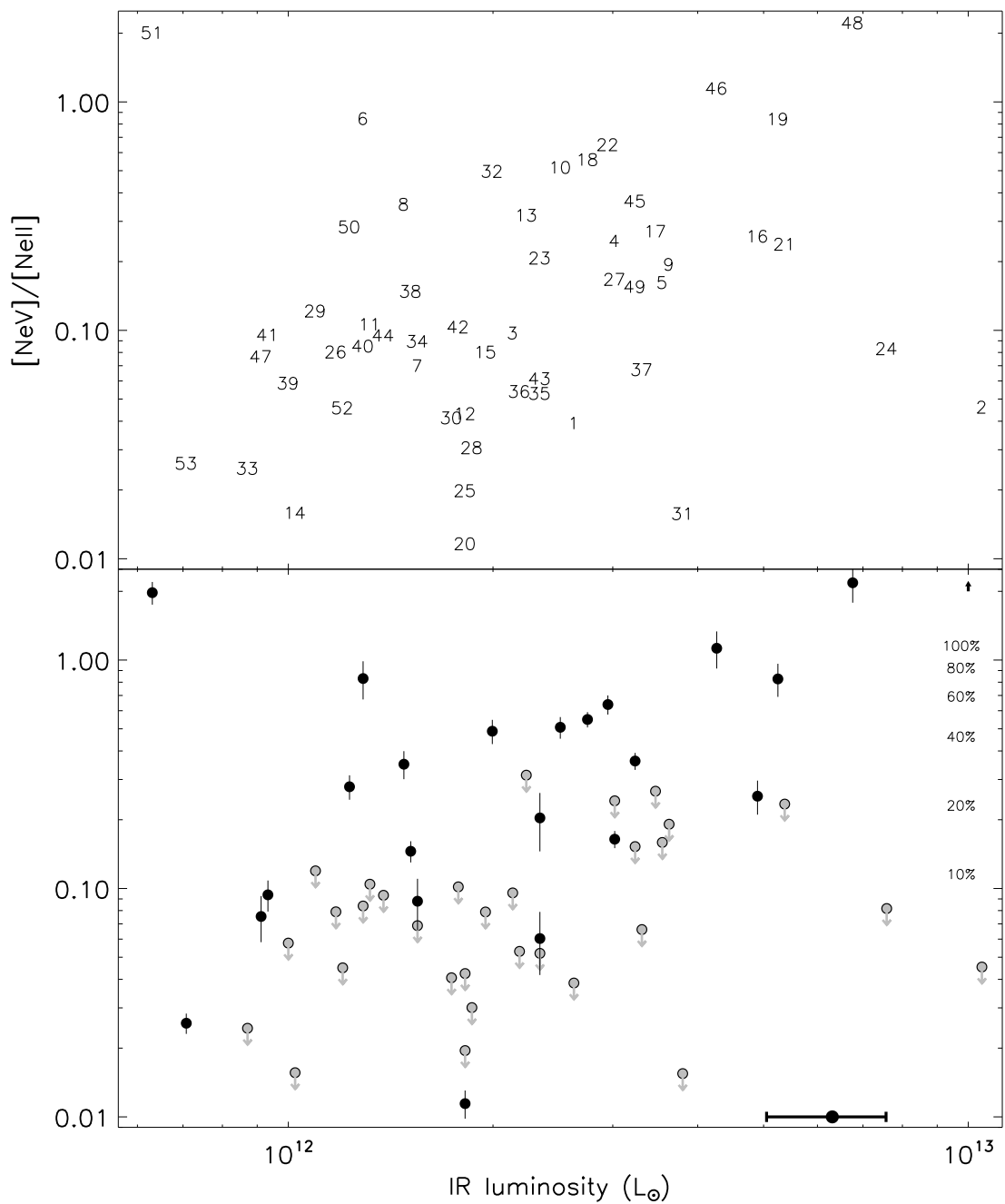}
\end{minipage}
\caption{[NeV]$\lambda14.32$/[NeII]$\lambda12.81$ vs infrared luminosity. The percentages on the $y$ axis indicate the predicted AGN 
contribution to the IR luminosity \citep{stu02}. ULIRGs with detections on both axes are plotted in black, while ULIRGs with limits 
on one or both axes are plotted in grey. The small arrow indicates the effect on a points position if the V band extinction 
towards the line emitting regions is increased by $A_{V}=30$. The horizontal bar on the bottom left indicates a 20\%\ error on the 
IR luminosity. \label{hiresdiaga}}
\end{figure}

\begin{figure}
\begin{minipage}{150mm}
\includegraphics[angle=90,width=150mm]{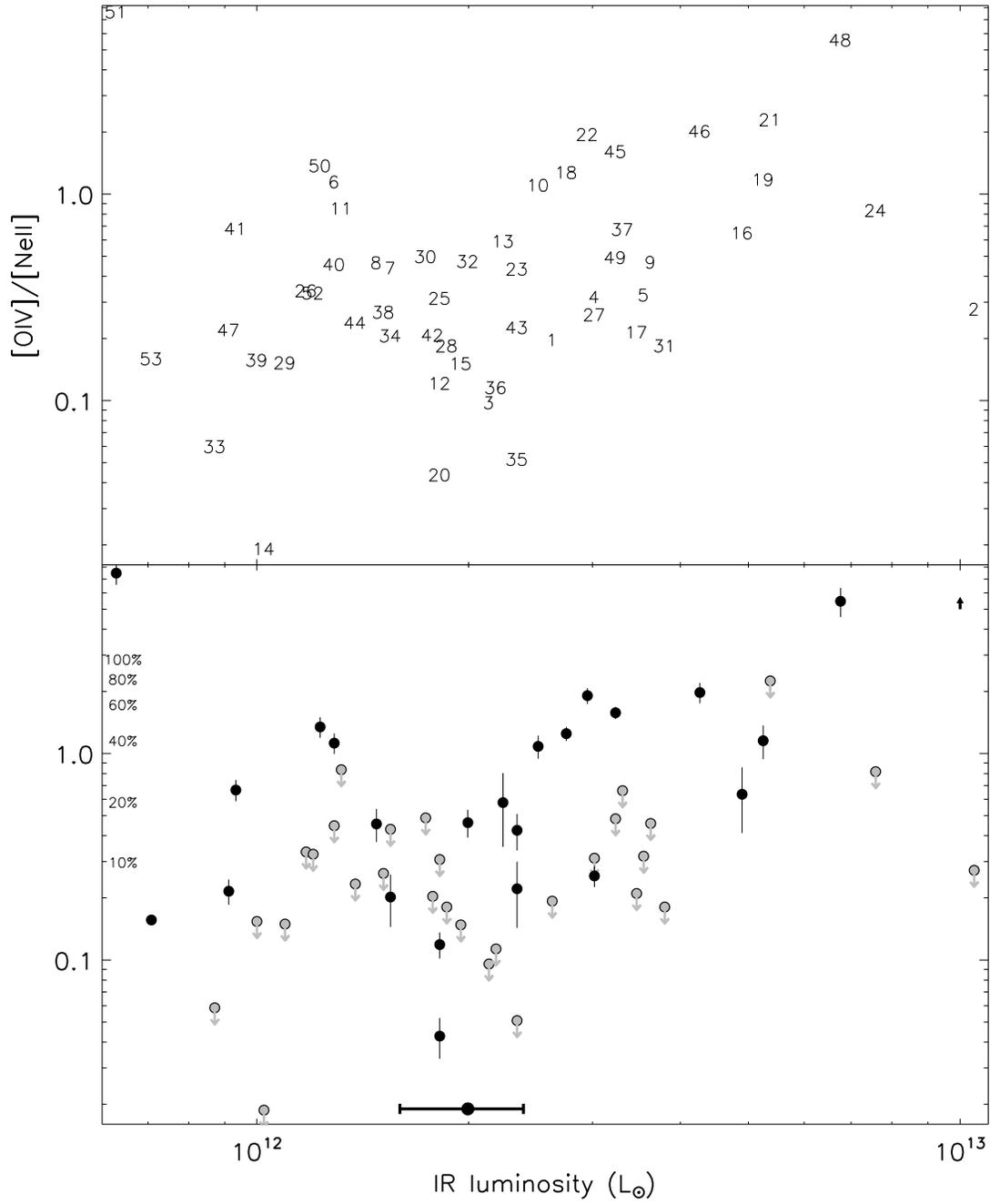}
\end{minipage}
\caption{[OIV]$\lambda25.89$/[NeII]$\lambda12.81$ vs infrared luminosity. Other data and symbols are the same as in Figure \ref{hiresdiaga}. \label{hiresdiagb}}
\end{figure}

\begin{figure}
\begin{minipage}{150mm}
\includegraphics[angle=90,width=150mm]{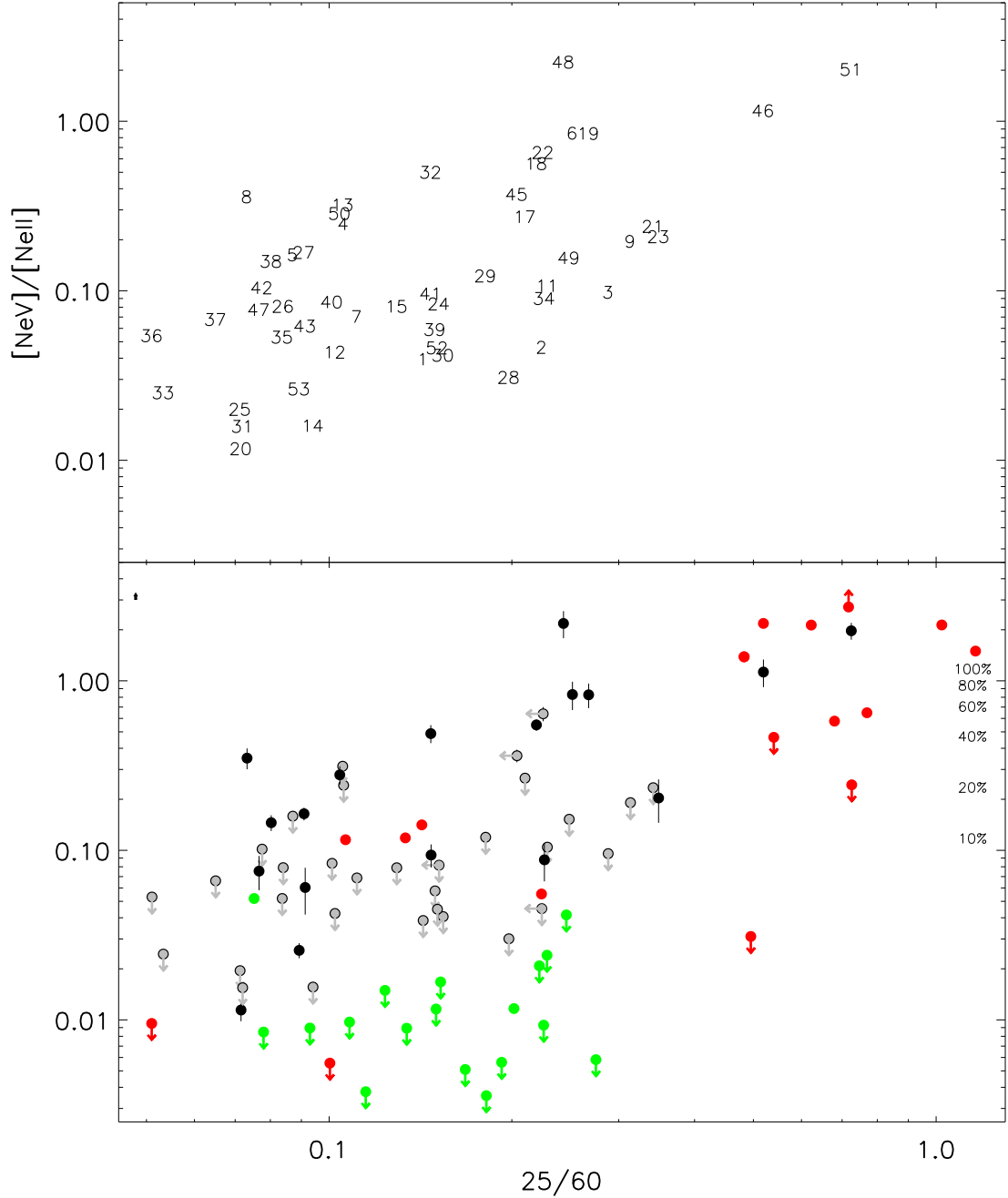}
\end{minipage}
\caption{[NeV]$\lambda25.89$/[NeII]$\lambda12.81$ vs IRAS 25/60 color. The green and red points are comparison starbursts and AGN respectively, but 
unlike Figure \ref{excite} these samples are taken from \citet{bra06} and \citet{wee05}, respectively. Other data and symbols are the same as in 
Figure \ref{hiresdiaga}.\label{nevcfwarm}}
\end{figure}

\begin{figure}
\begin{minipage}{170mm}
\includegraphics[angle=90,width=150mm]{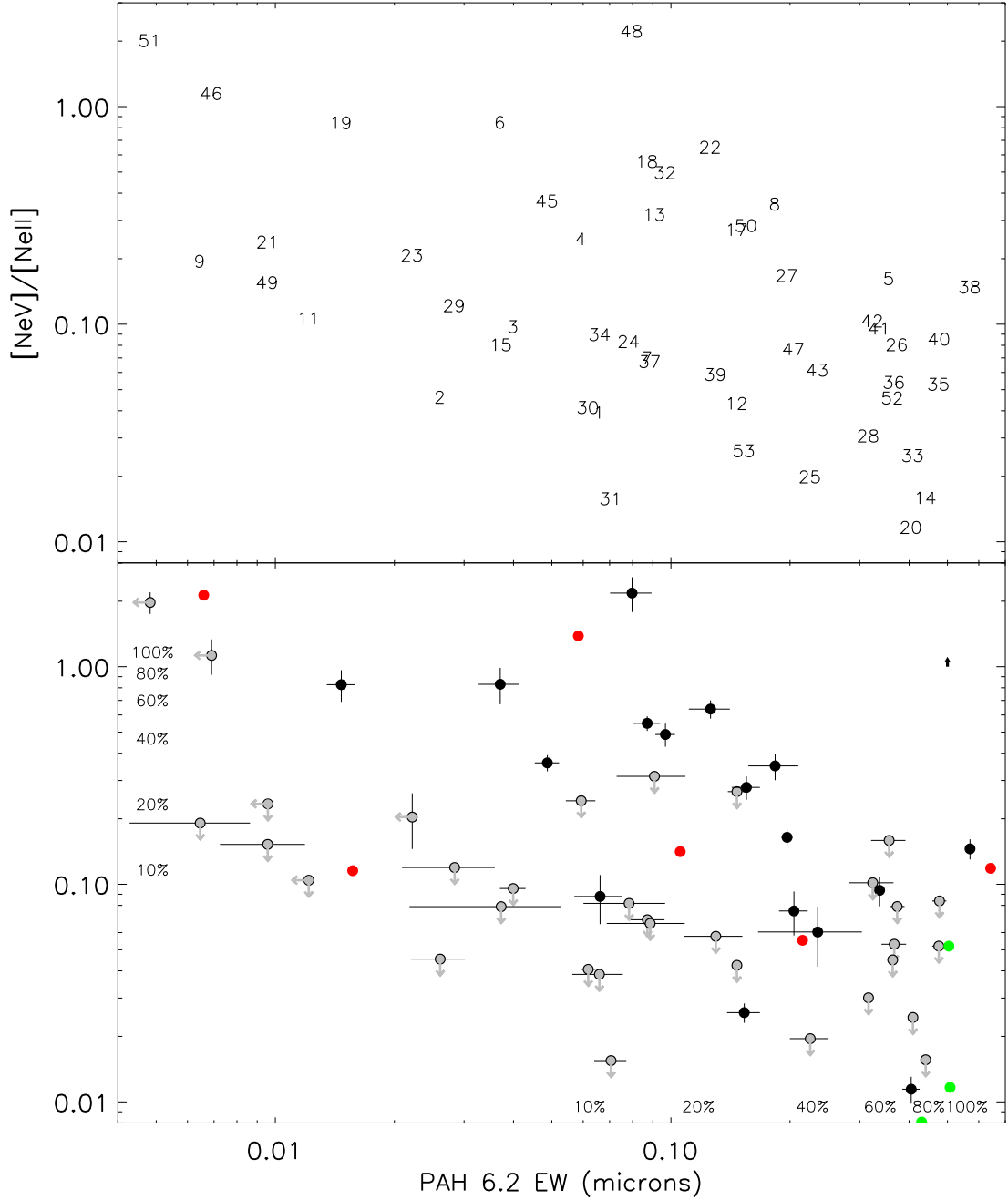}
\end{minipage}
\caption{[NeV]$\lambda14.32$/[NeII]$\lambda12.81$ vs the equivalent width of the PAH 6.2$\mu$m feature. The percentages on the $x$ axes indicate 
the predicted contribution to the IR luminosity from an starburst \citep{arm06}. Other data and symbols are the same as in 
Figure \ref{hiresdiaga}. \label{hipahdiaga}}
\end{figure}

\begin{figure}
\begin{minipage}{150mm}
\includegraphics[angle=90,width=150mm]{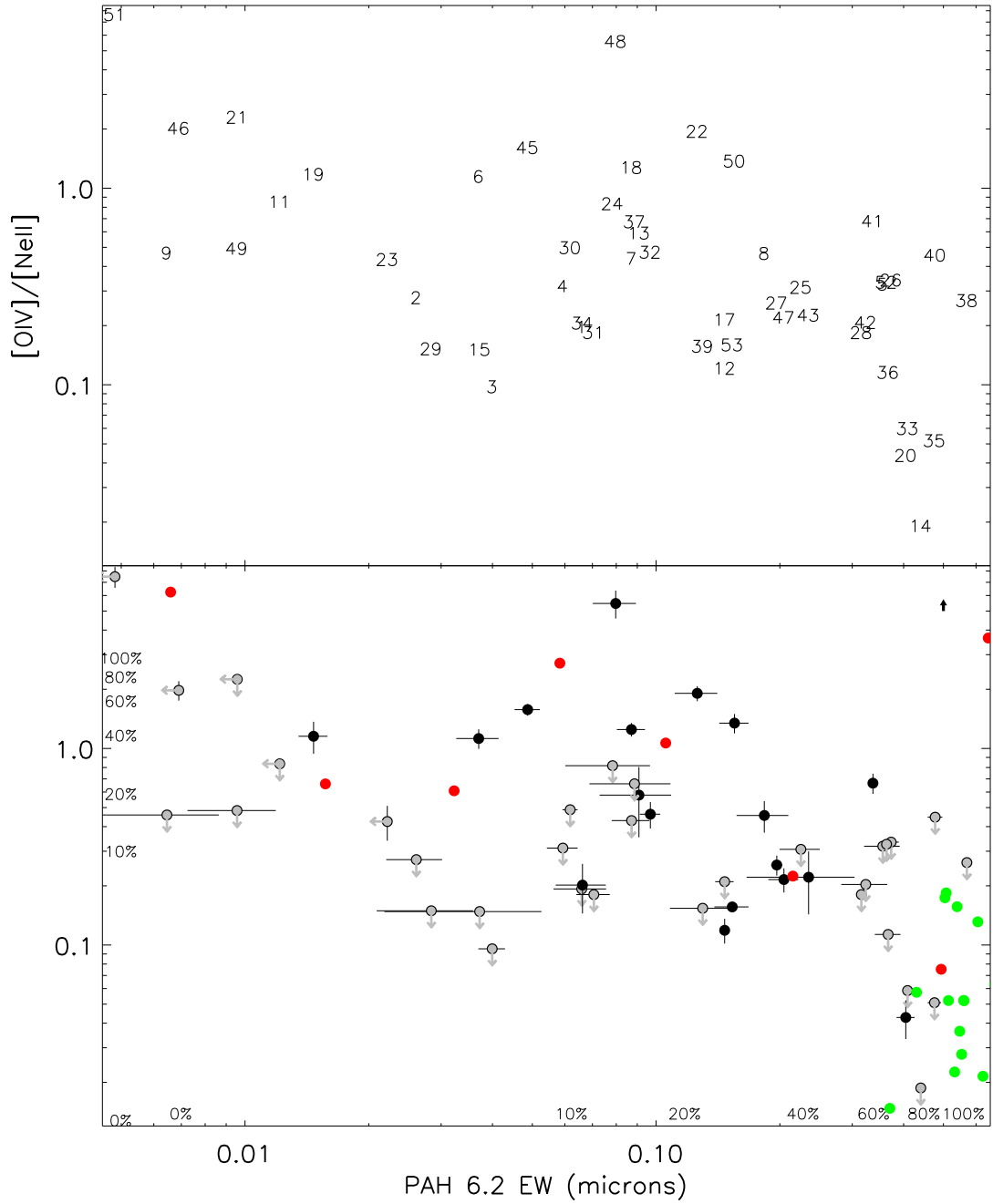}
\end{minipage}
\caption{[OIV]$\lambda25.89$/[NeII]$\lambda12.81$ vs the equivalent width of the PAH 6.2$\mu$m feature.  Other data and symbols are the same as in 
Figure \ref{hipahdiaga}. \label{hipahdiagb}}
\end{figure}

\begin{figure}
\begin{minipage}{160mm}
\includegraphics[angle=90,width=150mm]{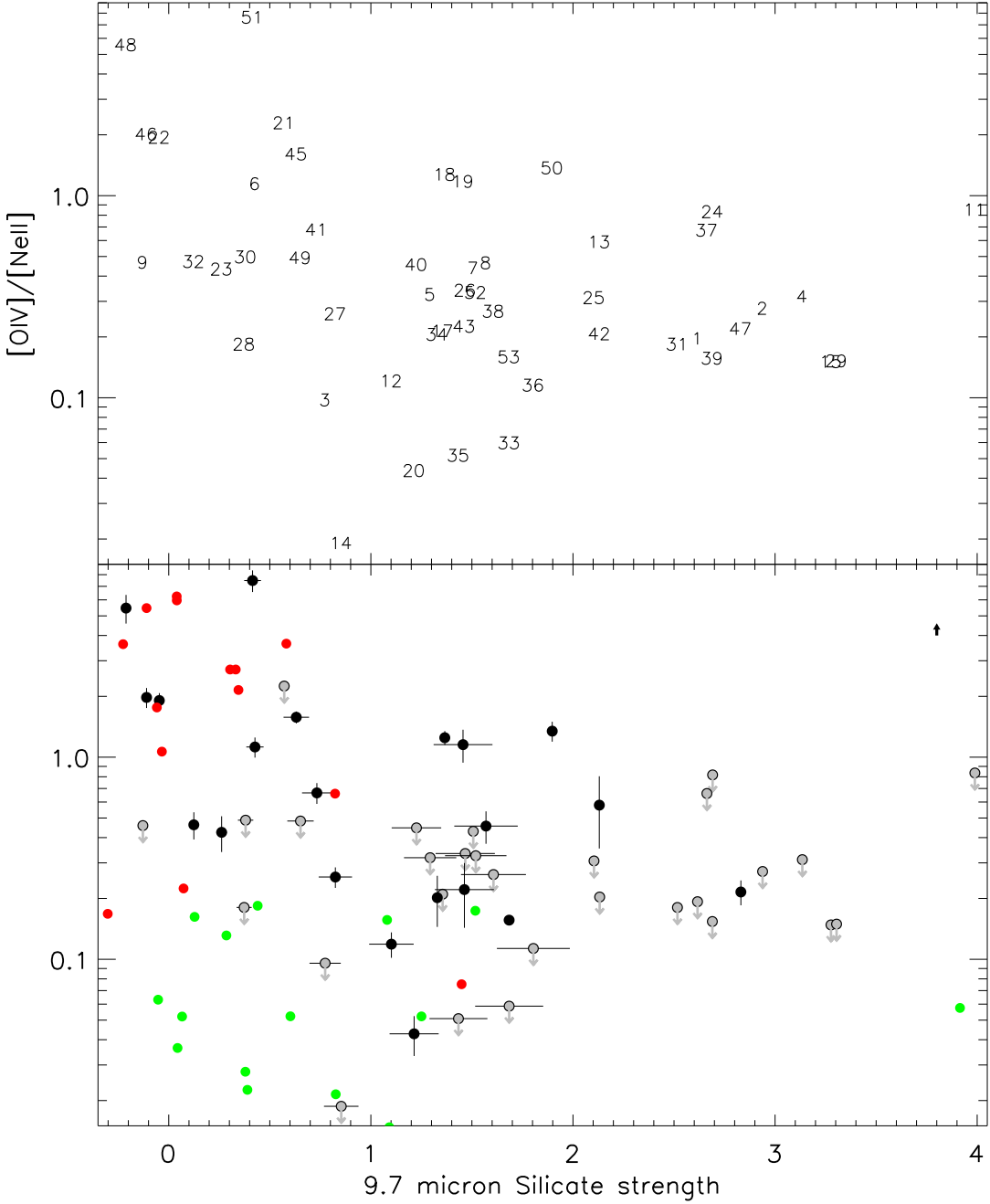}
\end{minipage}
\caption{[OIV]$\lambda25.89$/[NeII]$\lambda12.81$ vs the strength of the 9.7$\mu$m silicate feature. Other data and symbols are the same as in 
Figure \ref{hipahdiaga}. \label{oivsi}}
\end{figure}

\begin{figure}
\begin{minipage}{180mm}
\includegraphics[angle=90,width=170mm]{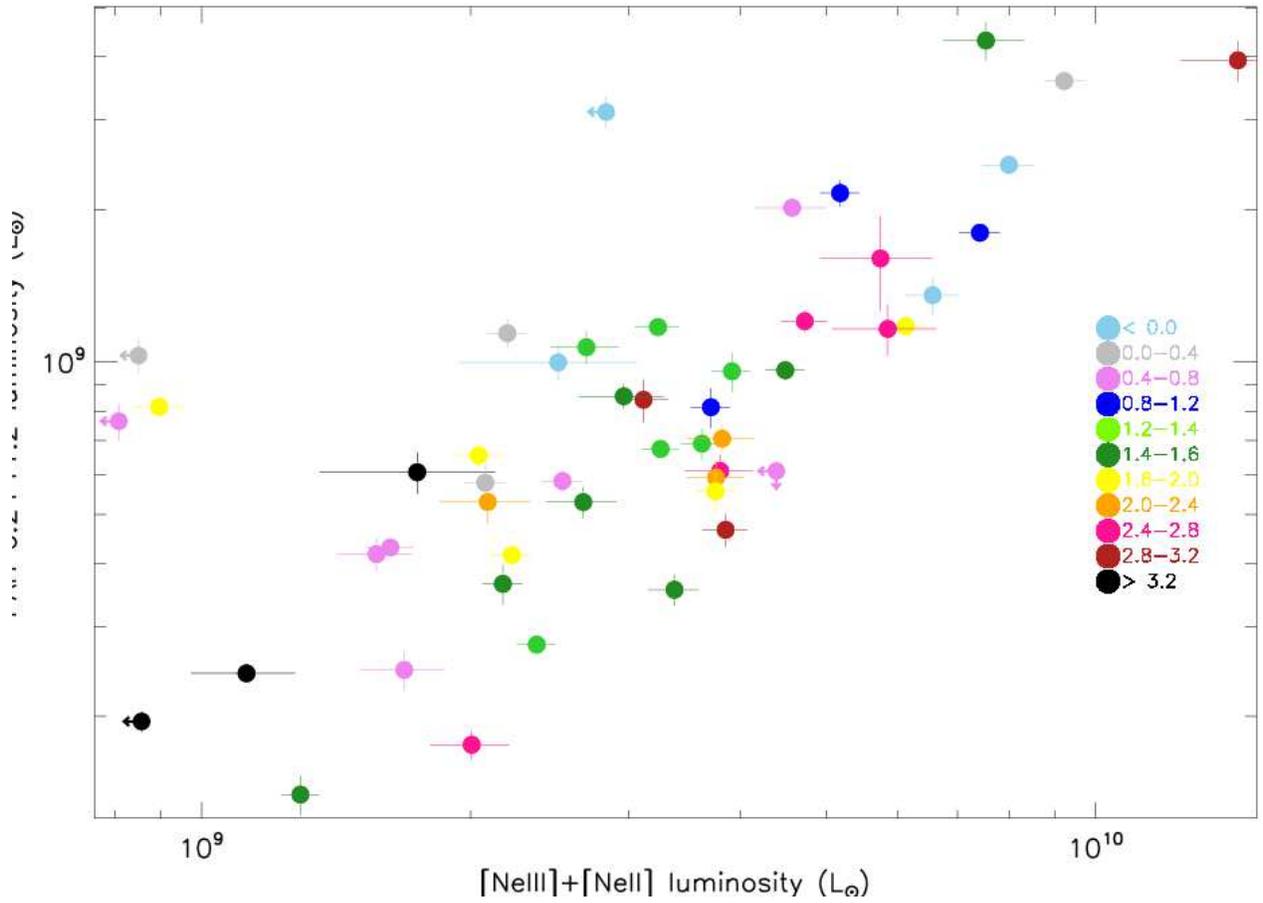}
\end{minipage}
\caption{The total luminosity of the [NeIII]$\lambda$15.55 and [NeII]$\lambda$12.81 lines vs the total luminosity of the PAH 6.2$\mu$m 
and 11.2$\mu$m features. The points are color coded according to the strength of the 9.7$\mu$m silicate feature. 
\label{sidepthcode}}
\end{figure}

\begin{figure}
\begin{minipage}{180mm}
\includegraphics[angle=90,width=170mm]{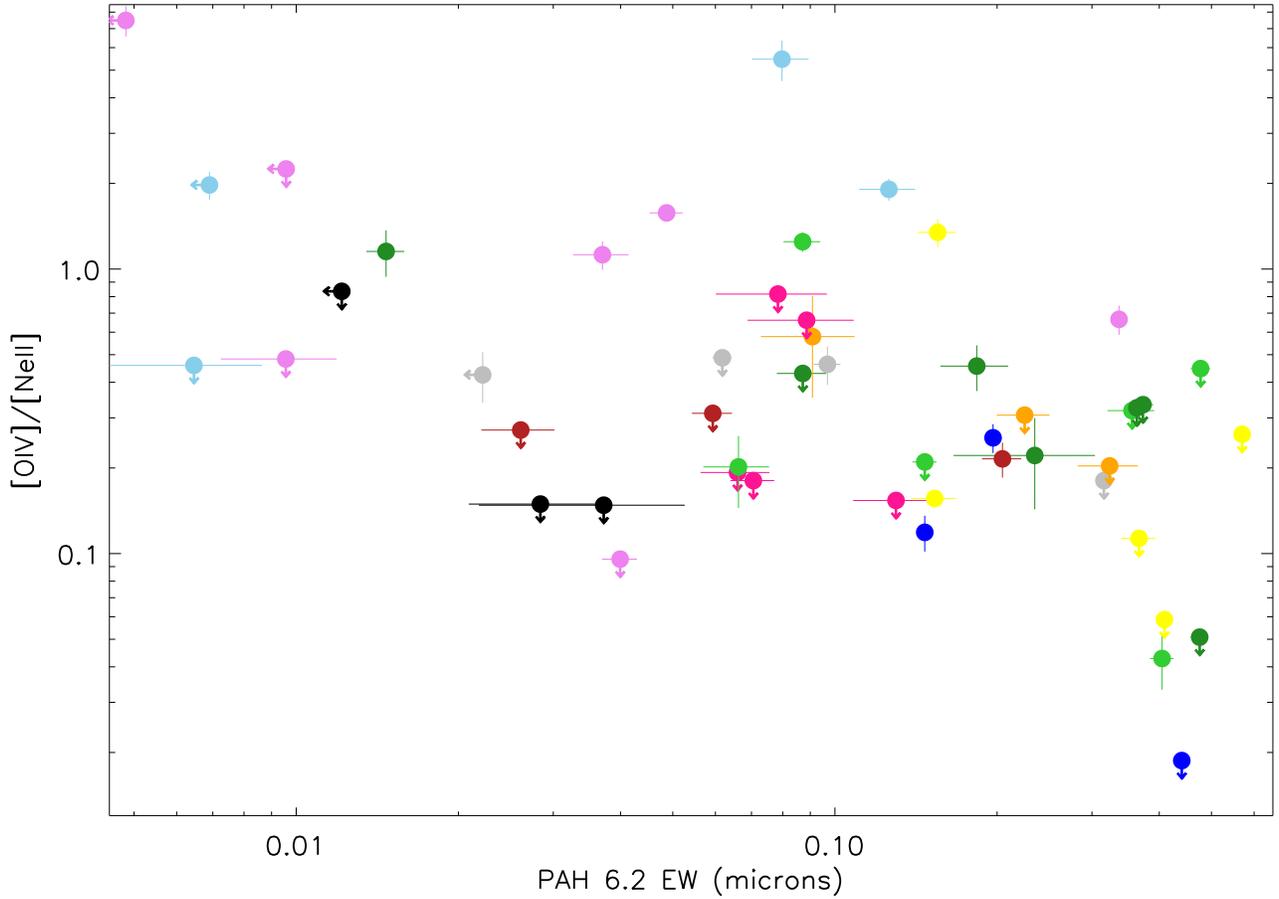}
\end{minipage}
\caption{[OIV]$\lambda25.89$/[NeII]$\lambda12.81$ vs the equivalent width of the PAH 6.2$\mu$m feature, with the points coded according to 
the strength of the 9.7$\mu$m silicate feature. The color coding is the same as in Figure \ref{sidepthcode}. Here we see that the 
lightly obscured and heavily obscured systems lie mostly on the left hand side of the plot, whereas moderately obscured 
systems lie mainly toward the right hand side of the plot. 
\label{sidepthcodeb}}
\end{figure}

\noindent 4) We show that the combined luminosity of the [NeIII]$\lambda$15.56 and [NeII]$\lambda$12.81 lines correlates with both total IR luminosity (Figure \ref{hoketo}), 
and the combined luminosity of the PAH 6.2$\mu$m and 11.2$\mu$m features (Figure \ref{hoketoalt}). Combining this result with previous work \citep{ho07}, we derive a 
calibration between star formation rate and PAH 6.2$\mu$m + 11.2$\mu$m luminosity for ULIRGs:

\begin{equation}
SFR [M_{\odot} yr^{-1}] = 1.18\times10^{-41}L_{P} [ergs\ s^{-1}]\label{pahstarconc}
\end{equation}

\noindent where $\stackrel{.}{M_{\odot}}$ is the star formation rate in solar masses per year, and $L_{P}$ is the PAH 6.2$\mu$m + 11.2$\mu$m luminosity 
in ergs s$^{-1}$

\noindent 5) We employ a variety of spectral diagnostics to show that, despite the presence of a luminous AGN in $\sim$42\% of ULIRGs, the 
most likely dominant contributor to the total IR emission in most ULIRGs is star formation, with an AGN providing a higher contribution 
than a starburst in only $\sim20\%$ of ULIRGs. The fine structure line ratios, luminosities and PAH EWs of our sample are consistent 
with the starbursts and AGN in ULIRGs being more extincted ($A_{V}\simeq40$), and for the starbursts more compact, versions of those in lower luminosity systems. 

\noindent 6) We show that the strength of the 9.7$\mu$m silicate feature is unlikely to be a simple indicator of the total obscuration of the starburst. 
We combine measurements of PAH equivalent widths, 9.7$\mu$m silicate feature strengths, and fine structure line ratios to show that ULIRGs with silicate 
strengths of $S_{sil}<0.8$ likely contain an energetically significant AGN, whereas the IR emission from ULIRGs with $0.8<S_{sil}<2.4$ is likely 
dominated by star formation. We postulate that ULIRGs with $S_{sil}>2.4$ contain an deeply buried AGN, 
though a comparably obscured starburst is also possible.

\acknowledgments
We thank Xander Tielens and Javier Goicoechea for advice on OH$^{-}$ absorption features, and the referee for a very helpful report. 
This work is based on observations made with the Spitzer Space Telescope, which is operated by the Jet Propulsion 
Laboratory, California Institute of Technology under a contract with NASA. Support for this work was provided by NASA.
This research has made extensive use of the NASA/IPAC Extragalactic Database (NED) which is operated by the Jet Propulsion 
Laboratory, California Institute of Technology, under contract with NASA.

\clearpage

\begin{deluxetable}{clcccccc}
\tabletypesize{\scriptsize}
\tablecolumns{8}
\tablewidth{0pc}
\tablecaption{Observations summary \label{sample}}
\tablehead{
\colhead{ID}&\colhead{Galaxy}&\colhead{RA (J2000)}&\colhead{Dec}&\colhead{Redshift}&\colhead{$L_{ir}$\tablenotemark{a}}&\colhead{$L_{radio}\tablenotemark{b}$}&\colhead{AOR Key\tablenotemark{c}}
}
\startdata
 1  & IRAS 00188-0856 & 00 21 26.5         & -08 39 26.3 & 0.128                  & 12.42   & 23.82  & 4962560   \\
 2  & IRAS 00397-1312 & 00 42 15.5         & -12 56 02.8 & 0.262                  & 13.02   & 23.67  & 4963584   \\
 3  & IRAS 01003-2238 & 01 02 50.0         & -22 21 57.5 & 0.118                  & 12.33   & 23.63  & 4964608   \\
 4  & IRAS 03158+4227 & 03 19 12.4         & +42 38 28.0 & 0.134                  & 12.48   & 23.74  & 12256256  \\
 5  & IRAS 03521+0028 & 03 54 42.1         & +00 37 03.4 & 0.152                  & 12.55   & 23.56  & 4968448   \\
 6  & IRAS 05189-2524 & 05 21 01.5         & -25 21 45.4 & 0.043                  & 12.11   & 23.06  & 4969216   \\
 7  & IRAS 06035-7102 & 06 02 54.0         & -71 03 10.2 & 0.079                  & 12.19   & --     & 4969728   \\
 8  & IRAS 06206-6315 & 06 21 01.2         & -63 17 23.5 & 0.092                  & 12.17   & --     & 4969984   \\
 9  & IRAS 07598+6508 & 08 04 33.1         & +64 59 48.6 & 0.148                  & 12.56   & 24.31  & 4971008   \\
 10 & IRAS 08311-2459 & 08 33 20.6         & -25 09 33.7 & 0.100                  & 12.40   & 24.21  & 4971520   \\
 11 & IRAS 08572+3915 & 09 00 25.4         & +39 03 54.4 & 0.058                  & 12.12   & 22.51  & 4972032   \\
 12 & IRAS 09022-3615 & 09 04 12.7         & -36 27 01.1 & 0.060                  & 12.26   & 23.84  & 4972288   \\
 13 & IRAS 10378+1109 & 10 40 29.2         & +10 53 18.3 & 0.136                  & 12.35   & 23.56  & 4974336   \\
 14 & IRAS 10565+2448 & 10 59 18.1         & +24 32 34.3 & 0.043                  & 12.01   & 23.37  & 4974848   \\
 15 & IRAS 11095-0238 & 11 12 03.4         & +02 04 22.4 & 0.107                  & 12.29   & 23.81  & 4975360   \\
 16 & IRAS 11119+3257 & 11 14 38.9         & +32 41 33.3 & 0.189                  & 12.69   & 24.96  & 4975616   \\
 17 & IRAS 12018+1941 & 12 04 24.5         & +19 25 10.3 & 0.169                  & 12.54   & 23.61  & 4976640   \\
 18 & IRAS 12071-0444 & 12 09 45.1         & -05 01 13.9 & 0.128                  & 12.44   & 23.48  & 4977408   \\
 19 & IRAS 12514+1027 & 12 54 00.8         & +10 11 12.4 & 0.319\tablenotemark{d} & 12.72   & 24.33  & 4978432   \\
 20 & IRAS 13120-5453 & 13 15 06.4         & -55 09 22.7 & 0.031                  & 12.26   & --     & 4978944   \\
 21 & IRAS 13218+0552 & 13 24 19.9         & +05 37 04.7 & 0.205                  & 12.73   & 23.69  & 4979200   \\
 22 & IRAS 13342+3932 & 13 36 24.1         & +39 17 31.1 & 0.179                  & 12.47   & 23.65  & 4979456   \\
 23 & IRAS 13451+1232 & 13 47 33.3         & +12 17 24.2 & 0.121                  & 12.37   & 26.26  & 4980480   \\
 24 & IRAS 14070+0525 & 14 09 31.3         & +05 11 31.8 & 0.264                  & 12.88   & 23.91  & 4980992   \\
 25 & IRAS 14348-1447 & 14 37 38.4         & -15 00 22.8 & 0.083                  & 12.26   & 23.75  & 4981248   \\
 26 & IRAS 14378-3651 & 14 40 59.0         & -37 04 32.0 & 0.068                  & 12.07   & 23.50  & 4981504   \\
 27 & IRAS 15001+1433 & 15 02 31.9         & +14 21 35.1 & 0.163                  & 12.48   & 23.99  & 4982272   \\
 28 & IRAS 15206+3342 & 15 22 38.0         & +33 31 35.9 & 0.124                  & 12.27   & 23.58  & 4982784   \\ 
 29 & IRAS 15250+3609 & 15 26 59.4         & +35 58 37.5 & 0.055                  & 12.04   & 22.95  & 4983040   \\
 30 & IRAS 15462-0450 & 15 48 56.8         & -04 59 33.6 & 0.100                  & 12.24   & 23.47  & 4984064   \\
 31 & IRAS 16090-0139 & 16 11 40.5         & -01 47 05.6 & 0.134                  & 12.58   & 23.96  & 4984576   \\
 32 & IRAS 17179+5444 & 17 18 54.2         & +54 41 47.3 & 0.147                  & 12.30   & 25.21  & 4986368   \\
 33 & IRAS 17208-0014 & 17 23 22.0         & -00 17 00.9 & 0.043                  & 11.94   & 23.63  & 4986624   \\
 34 & IRAS 19254-7245 & 19 31 21.6         & -72 39 22.0 & 0.063                  & 12.19   & 24.26  & 12256512  \\
 35 & IRAS 19297-0406 & 19 32 21.3         & -03 59 56.3 & 0.086                  & 12.37   & 23.64  & 4988672   \\
 36 & IRAS 20087-0308 & 20 11 23.9         & -02 59 50.7 & 0.106                  & 12.34   & 24.53  & 4989440   \\
 37 & IRAS 20100-4156 & 20 13 29.5         & -41 47 34.9 & 0.130                  & 12.52   & 23.90  & 4989696   \\
 38 & IRAS 20414-1651 & 20 44 18.2         & -16 40 16.2 & 0.087                  & 12.18   & 23.61  & 4989952   \\
 39 & IRAS 20551-4250 & 20 58 26.8         & -42 39 00.3 & 0.043                  & 12.00   & 23.11  & 4990208   \\
 40 & IRAS 22491-1808 & 22 51 49.3         & -17 52 23.5 & 0.078                  & 12.11   & 22.91  & 4990976   \\
 41 & IRAS 23128-5919 & 23 15 46.8         & -59 03 15.6 & 0.045                  & 11.97   & --     & 4991744   \\
 42 & IRAS 23230-6926 & 23 26 03.6         & -69 10 18.8 & 0.106                  & 12.25   & --     & 4992000   \\
 43 & IRAS 23253-5415 & 23 28 06.1         & -53 58 31.0 & 0.130                  & 12.37   & --     & 4992256   \\
 44 & IRAS 23365+3604 & 23 39 01.3         & +36 21 08.7 & 0.064                  & 12.14   & 23.36  & 4992512   \\
 45 & IRAS 23498+2423 & 23 52 26.0         & +24 40 16.7 & 0.212                  & 12.51   & 23.85  & 4992768   \\
 46 & Mrk 1014        & 01 59 50.2         & +00 23 40.6 & 0.163                  & 12.63   & 24.22  & 4966144   \\
 47 & UGC 5101        & 09 35 51.7         & +61 21 11.3 & 0.039                  & 11.96   & 23.73  & 4973056   \\
 48 & 3C 273          & 12 29 06.7         & +02 03 08.6 & 0.158                  & 12.83   & 27.47  & 4978176   \\
 49 & Mrk 231         & 12 56 14.2         & +56 52 25.2 & 0.042                  & 12.51   & 24.08  & 4978688   \\
 50 & Mrk 273         & 13 44 42.1         & +55 53 12.7 & 0.038                  & 12.09   & 23.63  & 4980224   \\
 51 & Mrk 463         & 13 56 02.9         & +18 22 19.1 & 0.050                  & 11.80   & 24.33  & 4980736   \\
 52 & Arp 220         & 15 34 57.1         & +23 30 11.5 & 0.018                  & 12.08   & 23.34  & 4983808   \\
 53 & NGC 6240        & 16 52 58.9         & +02 24 03.4 & 0.024                  & 11.85   & 23.92  & 4985600   \\
\enddata  
\tablenotetext{a}{Infrared luminosities are either taken from \citet{far4} or calculated using the formula in \citet{san96}, and converted 
to our cosmology. Units are the logarithm of the $1-1000\mu$m luminosity, in Solar luminosities (3.826$\times10^{26}$ W).}
\tablenotetext{b}{Observed-frame 1.4GHz luminosities, computed from the National Radio Astronomy Observatory Very Large Array Sky Survey (NVSS) catalogs \citep{con98}, in 
units of log (W). The six objects without luminosities are not within the NVSS survey area.}
\tablenotetext{c}{Astronomical Observing Request Key number.}
\tablenotetext{d}{Redshift is taken from the IRAS Point Source Catalog Redshift Survey (PSCz, \citealt{sau00}) rather than from \citet{wil98} as our spectra are consistent 
with the PSCz redshift.}
\end{deluxetable}

\clearpage

\begin{deluxetable}{lccccc}
\tabletypesize{\scriptsize}
\tablecolumns{6}
\tablewidth{0pc}
\tablecaption{Line ratio scaling factors for different extinction laws \label{extinctlaws}}
\tablehead{
\colhead{Line ratio}                                       &\colhead{Fluks\tablenotemark{a}}&\colhead{Li \& Draine\tablenotemark{b}}&\colhead{Draine\tablenotemark{c}}&\colhead{Chiar \& Tielens A\tablenotemark{d}}&\colhead{Chiar \& Tielens B\tablenotemark{e}}
}
\startdata    
 $\mathrm{\frac{[NeIII]\lambda15.56}{[NeII]\lambda12.81}}$ & 1.075         &  1.052               &  1.120         & 0.881                      &   0.734 \\
 $\mathrm{\frac{[SIII]\lambda18.71}{[SIII]\lambda33.48}}$  & 0.684         &  0.775               &  0.634         & --                         &   0.360 \\
 $\mathrm{\frac{[SIV]\lambda10.51}{[SIII]\lambda18.71}}$   & 0.498         &  0.621               &  0.506         & 0.664                      &   0.373 \\
 $\mathrm{\frac{[SIV]\lambda10.51}{[SIII]\lambda33.48}}$   & 0.340         &  0.478               &  0.321         & --                         &   0.134 \\
 $\mathrm{\frac{[OIV]\lambda25.89}{[NeII]\lambda12.81}}$   & 1.219         &  1.149               &  1.231         & 1.148                      &   0.660 \\
 $\mathrm{\frac{[NeV]\lambda14.32}{[NeII]\lambda12.81}}$   & 1.163         &  1.111               &  1.202         & 0.973                      &   1.000 \\
 $\mathrm{\frac{[NeV]\lambda14.32}{[NeV]\lambda24.32}}$    & 1.000         &  0.990               &  1.023         & 0.968                      &   1.644 \\
\enddata 
\tablecomments{Scaling factors assume an increase in the V band extinction of $A_{V}=30$, in the form of a screen.}
\tablenotetext{a}{\citet{flu94}}
\tablenotetext{b}{\citet{li01}}
\tablenotetext{c}{\citet{dra89}}
\tablenotetext{d}{\citet{chi06}, ISM}
\tablenotetext{e}{\citet{chi06}, Galactic Center}
\end{deluxetable}

\clearpage

\begin{deluxetable}{lcccccccccccccc}
\tabletypesize{\scriptsize}
\rotate
\tablecolumns{15}
\tablewidth{0pc}
\tablecaption{Common emission lines \label{linefluxa}}
\tablehead{
\colhead{Galaxy}&\colhead{[ArIII] }&\colhead{H$_{2}$,S(3)}&\colhead{ [SIV] }&\colhead{H$_{2}$,S(2)}&\colhead{[NeII] }&\colhead{ [NeV]} &\colhead{ [NeIII]}&\colhead{H$_{2}$,S(1)}&\colhead{[SIII]}&\colhead{[NeV] }&\colhead{[OIV]}&\colhead{H$_{2}$,S(0)}&\colhead{[SIII]  }&\colhead{[SiII]  } \\
\colhead{$\lambda_{rest}$} ($\mu$m)&\colhead{8.991 }  &\colhead{9.662  }&\colhead{10.511 }&\colhead{ 12.275}&\colhead{12.814 }&\colhead{14.322 }&\colhead{15.555 } &\colhead{17.030}&\colhead{18.713}&\colhead{24.318}&\colhead{25.890}&\colhead{28.219 }&\colhead{33.481  }&\colhead{34.815  } \\
\colhead{$E_{ion}$ (eV)}               &\colhead{27.6  }  &\colhead{ -- }     &\colhead{34.8 }  &\colhead{ --}      &\colhead{21.6   }&\colhead{97.1   }&\colhead{41.0 }   &\colhead{   --   }&\colhead{23.3  }&\colhead{97.1  }&\colhead{54.9  }&\colhead{  --     }&\colhead{23.3    }&\colhead{8.2 } 
}
\startdata     
00188-0856      & $<$0.15          &    0.75         & $<$0.26         &    0.32         &  4.67           &  $<$0.18        & 0.69             &    0.73        &    0.56::      & $<$1.60        & $<$0.90        &    1.39         & --               & --     \\
00397-1312      & 0.35             &    0.32:        &    0.30         & $<$0.27         &  4.41           &  $<$0.20        & 2.72:            &    1.09::      &    1.80:       & $<$1.50        & $<$1.20        & $<$0.75         & --               & --      \\
01003-2238      & --               &    0.77         &    0.21         &    0.51:        &  3.14           &  $<$0.30        & 1.31             &    1.08        & $<$0.90        & $<$0.30        & $<$0.30        & $<$1.00         & --                  & --                     \\ 
03158+4227      & $<$0.60          & $<$1.40         & $<$1.20         &    0.89         &  5.78           &  $<$1.10        & 0.94::           &    2.86:       &    1.71:       & $<$1.40        & $<$1.80        & $<$1.20         & --               & --                   \\
03521+0028      & $<$0.42          &    0.80         & $<$0.50         &    0.70         &  2.83           &  $<$0.45        & 1.27             &    1.94        &    0.87::      & $<$0.48        & $<$0.90        & $<$0.51         & --               & --               \\
05189-2524      & --               &    3.22         &    5.63         &    1.67         & 21.12           &    17.53        & 17.76   	        &    3.15        &    3.18        &   11.73        &   23.71        & $<$2.40         & $<$24.00         & 11.85::              \\ 
06035-7102      & --               &    2.05         & $<$0.39         &    1.20         &  6.98           &  $<$0.48        & 1.75:  	        &    2.71        &    3.00        & $<$0.81        &    $<$3.00      &    2.25         &     4.61:        & --                      \\
06206-6315      & --               &    0.73         &    0.21         &    0.50:        &  6.58           &     2.30        & 2.86             &    1.36        &    2.06        &    2.03        &    3.00        & $<$1.50         &     3.53         & --                \\
07598+6508      & $<$0.60          & $<$3.00         & $<$0.66         &    0.78::       &  3.92           &  $<$0.75        & 2.45             &    1.79        & $<$1.90        & $<$3.00        & $<$1.80        & $<$3.20         & --               & --                 \\
08311-2459      & --               &    2.41         &    8.70:        &    1.68:        & 24.29           &    12.34        & 22.53            &    2.80        &   11.43        &    9.81        &   26.30        & $<$4.40         &    18.40:        & --                   \\
08572+3915      & --               &    0.43:        & $<$0.50         & $<$0.63         &  7.18           &  $<$0.75        & 1.99:            &    0.90        &    1.69::      & $<$5.40        & $<$6.00        & $<$2.50         & $<$12.00         & $<$25.00              \\
09022-3615      & --               &    6.32         &    5.05         &    6.02         & 56.58           &  $<$2.40        & 40.00            &   10.81        &   24.73        & $<$3.60        &    6.72        &   11.16         &    41.33         & 95.83                 \\
10378+1109      & $<$2.00          &    1.10         & $<$0.41         &    0.85         &  3.51           &  $<$1.10        & 0.60::           &    0.55::      & $<$3.90        & $<$2.40        &    2.03::      & $<$2.70         & --               & --              \\
10565+2448      & --               &    4.05         & $<$0.27         &    2.53         & 64.06           &  $<$1.00        & 7.57             &    6.42        &   13.23        & $<$0.90        & $<$1.20        &    1.85::       &    23.92         & 43.40             \\
11095-0238      & --               &    1.80         & $<$1.20         &    1.19         &  6.08           &  $<$0.48        & 1.89             &    2.63:       &    1.22:       & $<$1.80        & $<$0.90        & $<$4.00         & --               & --                \\
11119+3257      & $<$0.30          &    0.52         & $<$1.50         &    0.47::       &  2.97           &     0.75        & 1.98             & $<$2.60        & $<$1.80        & $<$0.90        &    1.88::      & $<$1.50         & --               & --           \\
12018+1941      & $<$0.20          &    0.55:        & $<$0.19         &    0.60:        &  3.00           &  $<$0.80        & 0.35             & $<$1.80        & $<$0.70        & $<$0.10        & $<$0.63        &    2.67:        & --               & --           \\
12071-0444      & 0.61             &    1.36         &    1.46         &    0.85         &  5.25           &     2.88        & 5.09             &    2.12        &    1.86:       &    3.71        &    6.55        & $<$2.00         & --               & --                 \\ 
12514+1027      & 0.68             &    0.27::       &    0.91         &    0.50         &  2.35           &     1.94        & 2.68:            &    1.14::      &    1.21::      &    1.67:       &    2.71:       &    --           & --               & --              \\
13120-5453      &  --              &    6.89         &    0.50:        &    5.50         &150.04           &     1.71        & 18.46            &    9.49        &   19.18        & $<$20.00       &    6.42:       & $<$5.00         &    60.64         & 107.13             \\
13218+0552      & $<$0.09          &    0.44:        & $<$0.20         &    0.25:        &  1.07:          &  $<$0.25        & $<$0.85          &    0.98:       & $<$1.60        & $<$3.40        & $<$2.40        & $<$4.00         & --               & --                    \\
13342+3932      & 0.73             &    0.40::       &    2.12         &    0.35         &  5.69           &     3.45        & 4.97             &    1.20        &    2.97        &    4.23        &   10.32        &    0.72:        & --               & --               \\
13451+1232      & 0.51             &    1.82         &    0.69::       &    1.19         &  5.03:          &     1.02:       & 5.11             &    2.46        &    1.23:       &  $<$2.1        &    2.14:       &    1.23::       & --               & --                \\
14070+0525      & $<$0.27          &    0.58         & $<$0.20         &    0.29::       &  1.84:          &  $<$0.15        & 0.58:            &    0.56:       & $<$0.66        &  $<$0.60       & $<$1.50        & $<$1.10         & --               & --                 \\
14348-1447      &  --              &    2.87         &    0.25::       &    1.86         & 10.76           &  $<$0.21        & 2.59             &    4.97:       &    4.09:       &  $<$1.50       & $<$3.30        & $<$4.20         &    3.79:         & --           \\
14378-3651      &  --              &    0.95         & $<$0.60         &    1.25         & 11.39           &  $<$0.90        & 1.20             &    1.39        &    2.50        &  $<$2.30       & $<$3.80        & $<$2.90         &    5.74:         & $<$25.00                \\
15001+1433      & 0.44:            &    0.60         &    0.42:        &    0.37:        &  6.85           &     1.12        & 2.61             &    1.50:       &    2.25        &     0.66::     &    1.75        & $<$1.10         & --               & --           \\
15206+3342      & 2.10             &    0.54         &    4.35         &    0.35:        &  13.29          &  $<$0.4         & 20.52            &    1.06:       &    8.63        & $<$1.50        & $<$2.40        & $<$2.00         & --               & --           \\
15250+3609      &  --              &    0.74         & $<$0.69         &    0.60:        & 10.05           &  $<$1.20        & 2.68             &    1.17        &    4.27        &  $<$2.40       & $<$1.50        & $<$1.20         &    6.75:         & $<$19.00            \\
15462-0450      &  --              &    0.97         & $<$0.25         &    0.41:        &  7.38           &  $<$0.30        & 1.38:            &    1.45        &    1.78:       &  $<$1.05       & $<$3.60        & $<$2.00         &    4.80:         & --             \\
16090-0139      & 0.68:            &    1.52         & $<$0.20         &    0.79         &  7.76           &  $<$0.12        & 1.98:            &    1.85        &    1.83:       &  $<$2.00       & $<$1.40        & $<$3.00         & --               & --              \\
17179+5444      & 0.44:            &    1.16         &    0.46         &    0.78         &  4.54           &     2.21        & 2.90             &    2.01:       &    1.35:       &     0.82       &    2.10        & $<$0.24         & --               & --                    \\
17208-0014      &  --              &    5.45         & $<$0.40         &    6.06         & 41.22           &  $<$1.00        & 8.12             &    9.75        &    7.29        &  $<$3.20       & $<$2.40        &    3.41:        &   17.80          & 38.87:             \\
19254-7245      &  --              &    3.05         & $<$1.20         &    1.82         & 31.48           &     2.77:       & 13.19            &    6.14        &    4.28        &  $<$1.60       &    6.35:       &    2.63:        &    9.07          & 56.80::               \\
19297-0406      &  --              &    2.54         & $<$0.39         &    1.71         & 17.69           &  $<$0.92        & 2.46             &    3.38        &    4.93:       &  $<$2.20       & $<$0.90        & $<$6.00         &    9.15:         & --                 \\
20087-0308      &  --              &    1.59         & $<$0.48         &    1.14         & 14.14           &  $<$0.75        & 1.64             &    2.63        &    2.62        &  $<$1.90       & $<$1.60        & $<$2.10         &   17.73          & --                       \\
20100-4156      & 0.46:            &    1.03         & $<$0.20         &    0.60         &  7.26           &  $<$0.48        & 2.78:            &    1.19        &    2.74        &  $<$1.30       & $<$4.80        &    1.82:        & --               & --              \\
20414-1651      &  --              &    1.44         & $<$0.36         &    0.55:        &  6.85           &     1.00        & 1.57             &    1.60        &    2.16        &  $<$1.50       & $<$1.80        & $<$2.00         &    7.99          & --                \\
20551-4250      &  --              &    5.66         & $<$0.39         &    3.36         & 13.01           &  $<$0.75        & 2.79             &    7.93        &    4.26        &  $<$1.50       & $<$2.00        &    5.05:        & $<$10.00         & 22.29                \\
22491-1808      &  --              &    0.92         & $<$0.40         &    0.87         &  5.37           &  $<$0.45        & 1.85             &    1.67        &    2.69        &  $<$0.90       & $<$2.40        & $<$4.50         &     4.75:        & --                  \\
23128-5919      &  --              &    1.55         &    4.43         &    1.02         & 27.29           &     2.56        & 20.44            &    3.12        &   23.34        &     2.96:      &   18.16        & $<$12.00        &    22.47         & 17.48                 \\
23230-6926      &  --              &    0.93         & $<$0.70         &    0.54         &  7.38           &  $<$0.75        & 1.96             &    2.08        &    1.77        &  $<$1.20       & $<$1.50        & $<$3.00         &   $<$24.00       & --             \\
23253-5415      & $<$2.10          &    2.45         &    1.08         &    1.30         &  5.47           &     0.33::      & 1.87             &    3.68        &    1.66        &     1.21       &    1.20::      &    2.14::       &  --              & --            \\
23365+3604      &  --              &    1.42         & $<$0.75         &    1.01:        &  8.57           &  $<$0.80        & 0.73             &    2.11        &    3.01:       &  $<$0.54       & $<$2.00        & $<$6.20         &    3.81::        & 41.64::              \\
23498+2423      & 0.31:            &    0.44:        &    1.14:        &    0.21         &  3.17           &     1.15        & 2.71             &    0.85:       &    1.66:       &     2.01       &    5.00        &     0.53        &  --              & --            \\
Mrk 1014        & 0.69             &    0.67         &    3.69         &    0.30         &  6.57           &     7.40        & 9.71             &    1.15:       & $<$1.50        &     4.96       &   12.97        &     1.61        &   --             & --          \\
UGC 5101        & --               &    2.48         &    0.91:        &    2.40         & 34.13           &     2.57        & 13.66            &    4.59        &    5.57        &     2.82:      &    7.35        &     1.17        &    15.46         &   32.07           \\
3C 273          & $<$1.50          & $<$0.75         &    3.04:        & $<$0.80         &  1.55           &     3.38        & 6.00             &    1.21:       &    1.53        &     2.94       &    8.47        &  $<$2.40        &  --              &   --                 \\
Mrk 231         & --               &    3.56         & $<$2.10         &    4.24:        & 19.67           &   $<$3.0        & 3.05:            &    9.17        & $<$4.00        & $<$18.00       & $<$9.50        & $<$20.00        &   $<$8.00        & 16.00::             \\
Mrk 273         & --               &    9.38         &    9.58         &    5.21         & 41.90           &    11.68        & 33.57            &    8.63        &   13.35        &    15.38       &   56.36        &  $<$9.00        &   42.56          &   14.66:              \\
Mrk 463         & --               &    3.89         &   29.86         &    1.19         & 9.25           &    18.25         & 40.78            &    3.48        &   15.85        &    19.93       &   69.17        &     0.95::      &   15.50          &   29.79              \\
Arp 220         & --               &    --           & $<$1.50         &   10.06         & 64.54           &   $<$2.9        & 7.80             &   19.15        &    5.44        & $<$14.00       &$<$21.00        & $<$26.00        &   75.08::        &   32.28          \\
NGC 6240        & --               &   65.63         &    2.68         &   34.95         &171.22           &     4.40        & 60.65            &   44.29        &   17.13        &  $<$5.70       &   26.75        &     6.21:       &   38.11          &   265.86         \\
\enddata 
\tablecomments{Flux units are $\times 10^{-21}$ W cm$^{-2}$. A '--' indicates the line is outside the bandpass. Errors are of the 
order 10\%\ or less, except for those fluxes marked with a ':' which are 20\%\, and those marked with a '::' which are 30\%\ and should be considered suspect. }
\end{deluxetable}

\clearpage

\begin{deluxetable}{lcccccccccc}
\tabletypesize{\scriptsize}
%\rotate
\tablecolumns{11}
\tablewidth{0pc}
\tablecaption{Unusual emission lines \label{linefluxc}}
\tablehead{
\colhead{Galaxy}                   &\colhead{HI 7-6}&\colhead{[ArV]} &\colhead{[ClII]}&\colhead{[PIII]}&\colhead{[FeII]}&\colhead{[FeIII]}&\colhead{[FeII]}&\colhead{[FeII]}&\colhead{[NeIII]}\\
\colhead{$\lambda_{rest}$} ($\mu$m)&\colhead{12.368}&\colhead{13.102}&\colhead{14.365}&\colhead{17.885}&\colhead{17.936}&\colhead{22.925 }&\colhead{24.519}&\colhead{25.988}&\colhead{36.014} \\
\colhead{$E_{ion}$ (eV)    }       &\colhead{7.9 }  &\colhead{59.8  }&\colhead{109.2 }&\colhead{13.0}  &\colhead{19.8 } &\colhead{7.9 }   &\colhead{16.2 } &\colhead{7.9 }  &\colhead{41.0 } 
}
\startdata    
01003-2238                  &  --            & 0.46           &   --           &  --            &  --            &  --             & --             &  --            &  --   \\
03521+0028                  &  --     	     & --             &   --           &  --            &  --            &  --             & 0.88           &  --            &  --   \\
09022-3615                  &  --     	     &  --            &   --           &  --            &  --            &  --             & --             &  5.28:         &  --   \\
10565+2448                  &  --            &  --            &  0.99          &  --            &  --            &  --             & --             &  1.69:         &  --   \\
12018+1941                  &  --            &  --            &  0.30::        &  --            &  --            &  --             & --             &  --            &  --   \\
12514+1027                  &  --     	     & --             &   --           & 1.35::         &  --            &  --             & --             &  --            &  --   \\
13120-5453                  &  0.75::        &  --            &    --          &  --            &  --            &  --             & --             &  5.37          &  --   \\
14070+0525                  &  0.43:         &  --            &    --          &  --            &  --            &  --             & --             &  --            &  --   \\
15206+3342                  &  0.23:         &  --            &    --          &  --            &  --            &  1.00           & --             &  --            &  --   \\
17208-0014                  &  --     	     &  --            &   --           &  --            &  --            &  --             & --             &  2.80::        &  --   \\
19297-0406                  &  --            &  --            &    --          &  --            &  --            &  --             & --             &  1.70          &  --  \\
20100-4156                  &  0.28:         &  --            &    --          &  --            &  --            &  --             & --             &  --            &  --   \\
23498+2423                  &  --            &  --            &   0.70         & 0.70:          &  --            &  --             & --             &  --            &  --  \\
Mrk 463                     &  --            & 0.97           &  --            &  --            &  --            &  --             & --             &  --            &  --  \\
UGC 5101                    &  --            &  --            &   0.79:        &  --            &  --            &  --             & --             &  3.14::        &  --  \\
NGC 6240                    &  --            &  --            &   2.38         &  --            &  5.49          &  --             & --             &  21.62         &  --  \\
Arp 220                     &  --            &  --            &   2.51         &  --            &  --            &  --             & --             &  --            & 48.03::\\
Mrk 231                     &  --            &  --            &  --            &  --            &  --            &  --             & --             &  --            &  --  \\
\enddata 
\tablecomments{Flux units are $\times 10^{-21}$ W cm$^{-2}$. In this table, a '--' indicates that the 
line is not seen, however we do not quote upper limits due to the uncertain nature of the line IDs. Formal errors are all of the 
order 10\%, except for ':' which are 20\%, and '::' which are 30\%. }
\end{deluxetable}

\clearpage

\clearpage

\clearpage

\clearpage

\clearpage

\clearpage

\clearpage

\clearpage

\clearpage

\clearpage

\clearpage

\clearpage

\clearpage

\clearpage

\clearpage

\clearpage

\clearpage

\clearpage

\clearpage

\clearpage

\clearpage

\clearpage

\clearpage

\end{document}